\documentclass{article}

\usepackage{arxiv}

\usepackage[utf8]{inputenc} % allow utf-8 input
\usepackage[T1]{fontenc}    % use 8-bit T1 fonts
\usepackage{hyperref}       % hyperlinks
\usepackage{url}            % simple URL typesetting
\usepackage{booktabs}       % professional-quality tables
\usepackage{amsfonts}       % blackboard math symbols
\usepackage{nicefrac}       % compact symbols for 1/2, etc.
\usepackage{microtype}      % microtypography
\usepackage{lipsum}		% Can be removed after putting your text content
\usepackage{graphicx}
\usepackage{fullpage}

\usepackage{amsmath,amssymb,amsfonts,amsthm}
\theoremstyle{definition}

\theoremstyle{lemma}

\theoremstyle{theorem}

\theoremstyle{assumption}

\usepackage[titletoc,toc,title]{appendix}
\usepackage{textpos}
\usepackage{array} 
\usepackage{listings}
\usepackage{mathtools}
\usepackage{pdfpages}
\usepackage[textsize=footnotesize,color=green]{todonotes}
\usepackage{bm}
\usepackage{bbm}

\usepackage{tikz}
\usepackage[normalem]{ulem}
\usepackage{hhline}

\usepackage{graphicx}
\usepackage{subfig}
\usepackage{color}

\usepackage{pgfplots}
\usepackage{pgfplotstable}
\definecolor{markercolor}{RGB}{124.9, 255, 160.65}
\pgfplotsset{
compat=1.3,
width=10cm,
tick label style={font=\small},
label style={font=\small},
legend style={font=\small}
}

\usetikzlibrary{calc}
\usetikzlibrary{intersections} 

\usepackage{stmaryrd}

\newcommand{\eval}[2][\right]{\relax
  \ifx#1\right\relax \left.\fi#2#1\rvert}

\newcolumntype{C}[1]{>{\centering\let\newline\\\arraybackslash\hspace{0pt}}m{#1}}

\makeatletter
\renewcommand\d[1]{\mspace{6mu}\mathrm{d}#1\@ifnextchar\d{\mspace{-3mu}}{}}
\makeatother

\graphicspath{{./figs/}}

\title{High-Order Methods for Hypersonic Flows with Strong Shocks and Real Chemistry}

\author{ {Ahmad Peyvan}\\
	Division of Applied Mathematics\\
	Brown University\\
	Providence, RI 02906\\
	\And
    {Khemraj Shukla} \\
	Division of Applied Mathematics\\
	Brown University\\
	Providence, RI 02906\\
	\And
    {Jesse Chan} \\
	Department of Computational and Applied Mathematics\\
	Rice University\\
        Houston, TX, 77054, United States\\
	\And
    {George Em Karniadakis}\thanks{Corresponding author.\textit{E-mail address:} george\_karniadakis@brown.edu}\\
    Division of Applied Mathematics\\
	Brown University\\
	Providence, RI 02906\\
}

\begin{document}
\maketitle

\begin{abstract}
% \label{df_sec}
%%%
%We present a new high-order scheme for the simulation of hypersonic flows. 
We compare high-order methods including spectral difference (SD), flux reconstruction (FR), the entropy-stable discontinuous Galerkin spectral element method (ES-DGSEM), modal discontinuous Galerkin methods, and WENO to select the best candidate to simulate strong shock waves characteristic of hypersonic flows. We consider several benchmarks, including the Leblanc and modified shock-density wave interaction problems that require robust stabilization and positivity-preserving properties for a successful flow realization. We also perform simulations of the three-species Sod problem with simplified chemistry with the chemical reaction source terms introduced in the Euler equations. The ES-DGSEM scheme exhibits the highest stability, negligible numerical oscillations, and requires the least computational effort in resolving reactive flow regimes with strong shock waves. Therefore, we extend the ES-DGSEM to hypersonic Euler equations by deriving a new set of two-point entropy conservative fluxes for a five-species gas model. 
Stabilization for capturing strong shock waves occurs by blending high-order entropy conservative fluxes with low-order finite volume fluxes constructed using the HLLC Riemann solver. The hypersonic Euler solver is verified using the non-equilibrium chemistry Sod problem. To this end, we adopt the Mutation++ library to compute the reaction source terms, thermodynamic properties, and transport coefficients. We also investigate the effect of real chemistry versus ideal chemistry, and the results demonstrate that the ideal chemistry assumption fails at high temperatures, hence real chemistry must be employed for accurate predictions. Finally, we consider a viscous hypersonic flow problem to verify the transport coefficients and reaction source terms determined by the Mutation++ library. 

% - High pressure ratio (Leblanc)?Which high-order method can handle without chemistry? compare DGSEM vs SD vs FR 
% leblanc case and modified Shu-Osher cases

% - Air dissociation with simplified chemistry(Shu three species problem) DGSEM vs SD vs FR vs WENO Shu's results vs DG of Shu (Pressure ratio of 1000)

% - Real chemistry with 5 species with mutaion++ library. (Validation of DGSEM with Grossman test case (sod), equilibrium vs non-equilibrium 

% - Real chemistry plus viscosity (Validation vs Marxen)
%%%%
\end{abstract}

% \begin{keyword}
% % MSC codes here, in the form: \MSC code \sep code
% % or \MSC[2008] code \sep code (2000 is the default)
% % \MSC 41A05\sep 41A10\sep 65D05\sep 65D17
% % Keywords
% \KWD High-order numerical schemes\sep  Discontinuous spectral element method\sep Entropy stable schemes \sep Finite rate chemical reaction \sep Hypersonic flows \sep Non-equilibrium chemistry \sep High pressure ratio shock tubes
% \end{keyword}

% \end{frontmatter}

%\linenumbers

%% main text
\section{Introduction}
\label{intro}
Accurate prediction of hypersonic flows plays a crucial role in the design of high-speed vehicles such as space capsules, hypersonic aircrafts, and projectiles. For example, the aerodynamic forces that act on a space capsule during re-entry directly affect the design of the navigation and control surfaces of the capsule. Calculating the thermal loads leads to the design of safe heat shield tiles with optimal thickness for the hypersonic regimes. Hypersonic flow simulations require a robust, accurate, and scalable numerical method to capture multiscale and chemically reactive compressible flows. The numerical method must resolve strong bow shocks with extreme pressure and temperature ratios, which trigger chemical reactions such as air dissociation and recombination. The chemical reactions progress with a finite rate, which violates the frozen chemical composition assumption and chemical equilibrium with infinite rate chemistry. Moreover, the turbulent boundary layer behind strong bow shocks must be resolved for a vast range of wavenumbers to precisely predict the surface forces and heat loads. A candidate numerical method is the high-order discontinuous spectral element methods (DSEM) that exhibits low numerical dissipation and dispersion errors \cite{peyvan2021flux} at high wavenumbers, which adds accuracy in predicting the turbulent boundary layer of hypersonic vehicles. These discontinuous high-order methods can be discretized onto local elements providing a framework suitable for parallel computations. However, DSEM lacks robustness when resolving strong shock waves with extreme pressure ratios. 

The majority of the existing numerical approaches for solving hypersonic flows consists of low-order finite volume methods with various Euler flux formulations \cite{kitamura2013towards,hao2016numerical}. Low-order finite volume methods exhibit a high degree of robustness. However, due to  excessive numerical errors, especially in long-time simulations, they encounter numerical issues such as bow shock carbuncle \cite{henderson2007grid,kitamura2010evaluation,pandolfi2001numerical} and the erroneous prediction of surface heating in hypersonic flows \cite{kitamura2010evaluation}. Predicting the surface heat fluxes is essential since it directly determines heat shield design for hypersonic re-entry. Finite volume methods are also dependent on the type of mesh elements and grid orientation \cite{gnoffo2007simulation}. Non-aligned grids with strong bow shocks induce numerical errors, leading to the incorrect prediction of surface heating \cite{gnoffo2007simulation,ching2019shock}. Grid alignment issue can be alleviated by adapting grid lines to the bow shock shape \cite{paciorri2009shock,bonfiglioli2013unstructured}; however, detecting the location of the bow shock creates implementation issues for complex geometries. Finite volume methods also fail to resolve the turbulent boundary layer accurately at high speeds. 

High-order methods may potentially overcome the deficiencies of low-order methods in simulating hypersonic flows. Marxen \emph{et al.}~\cite{marxen2013method} showed that a high-order finite difference scheme could successfully resolve a reactive turbulent boundary layer in equilibrium and non-equilibrium conditions for Mach number 10. Passiatore \emph{et al.}~\cite{passiatore2021finite} simulated the effect of finite rate chemistry in turbulent hypersonic boundary layers using a $10^\textrm{th}$ order finite difference scheme with localized artificial diffusion method added for numerical stability. Although the high-order finite difference scheme obtains promising results for the hypersonic boundary layer, it cannot be directly applied to deformed elements required in complex-geometry flows. Discontinuous spectral element methods provide the geometric flexibility of finite volume as well as high-order accuracy \cite{karniadakis2005spectral}. Ching \emph{et al.}~\cite{ching2019shock} developed a high-order discontinuous Galerkin (DG) methods for high Mach number flows. They showed that their DG method could resolve bow shocks without carbuncles and without a shock-aligned mesh. However, they did not employ real chemistry effects and assumed frozen properties for the flow.

A high-order scheme must be able to resolve non-equilibrium chemical reactions, strong shock waves, and near-vacuum conditions that are specific to hypersonic flows. Zhang and Shu \cite{zhang2010positivity} first introduced a high-order DG scheme that preserves the positivity of density and pressure near vacuum conditions for Euler equations. The DG scheme can simulate high-pressure ratio shock waves using a combination of total variation bounded (TVB) \cite{cockburn1989tvbs} method and a positivity preserving scheme that maintains high-order accuracy. They later extended the positivity preserving scheme to Euler equations with reaction source terms \cite{zhang2011positivity}. However, their method  failed to correctly predict the three species high-pressure ratio Sod problem with reaction source terms. Zhang and Shu \cite{zhang2012positivity} also developed a positivity preserving scheme for a fifth-order finite difference WENO scheme and successfully solved the three species Sod problem with reaction source terms. 

For collocation  discontinuous spectral element methods such as discontinuous Galerkin spectral element method (DGSEM), spectral difference (SD), and flux reconstruction (FR) approaches, a nonlinear modal filter can be employed to preserve the high accuracy, conservation, positivity of pressure and density, and entropy stability \cite{dzanic2022positivity}. The optimal modal filter strength is selected to satisfy the entropy stability and positivity of pressure and density for all the solution points using a bisection root-finding method. The adaptive modal filtering method is yet to be evaluated on multiple species reactive Euler and Navier-Stokes equations. Another viable stabilizing technique for high-pressure ratio shock waves includes entropy-stable high order DGSEM schemes for which entropy stability is recovered at the discrete level \cite{hennemann2021provably}. Constructing entropy stable DG approaches requires the derivation of the entropy conservative fluxes \cite{gassner2016split,chan2018discretely}. The entropy conservative fluxes have been derived for multi-component gas mixtures \cite{gouasmi2020formulation} when the specific heat coefficients are constant or a  polynomial function of the temperature. However, entropy conservative fluxes have yet to be derived for hypersonic flows, where the internal energy consists of the translational, rotational, and vibrational energies. In the current study, we develop a new set of entropy-conservative fluxes that conserve entropy for hypersonic flows.

Air is a gas mixture consisting of atomic and diatomic species. At high speeds, bow shocks generate hot gas regions behind the shocks that excite the vibrational and electronic modes of the gas mixture. At high temperatures, the thermodynamic and transport properties of the gas mixture are temperature-dependent. Moreover, when the temperature exceeds a certain limit, the dissociation and recombination reaction rates begin to increase, the composition of the gas mixture changes, and the frozen gas mixture assumption fails. The gas mixture can be assumed in local equilibrium when the advection characteristic time is much larger than the reaction time scale. For hypersonic flows, the advection  time scale is comparable to the reaction time scale due to high velocities. A small advection time scale must be modeled with non-equilibrium chemistry. At even higher velocities, ionization of the gas mixture begins generating species with electric charges such as electrons and ions. The air has been modeled as three- \cite{eberhardt1987shock}, five- \cite{marxen2013method}, and eleven-species \cite{Gupta1990} gas mixtures. In this work, the five-species model of Park \cite{park1993review} is employed for finite rate chemistry and non-equilibrium thermodynamic calculations. 

In this current study we aim to develop a high-order robust numerical framework for simulating hypersonic flows modeled with real chemistry. The high-order framework is constructed based on the DGSEM approach stabilized using a flux blending technique \cite{hennemann2021provably}. The blending is constructed such that the high accuracy of the DGSEM scheme is preserved for the smooth regions, and the conservation property remains intact. The DGSEM scheme is reformulated as a flux differencing expression, where the two points entropy conservative (EC) fluxes can be employed. The novel elements of this work include:

\begin{itemize}
    \item Comparing existing state-of-the-art discontinuous spectral element methods for resolving extreme pressure ratio shock waves in near vacuum conditions.
    \item Constructing an entropy-stable high order scheme for the hypersonic Euler equations.
    \item Deriving entropy-conservative fluxes for the formulation of hypersonic flows modeled using the Rigid-Rotor Harmonic-Oscillator (RRHO) method \cite{Scoggins2020}.
    \item Deriving a new technique to approximate an averaging term in the formulation of the two-point entropy-conservative (EC) fluxes when the left and right states approach the same values.
    \item Quantifying the effect of modeling real chemistry in hypersonic flows versus ideal chemistry.
    \item Constructing a hypersonic Navier-Stokes solver based on the entropy-stable Euler scheme and the Bassi and Rebay (BR1) approach \cite{bassi1997high} for discretizing the viscous terms.
    \item Validating the Mutation++ library for calculating the species source term, thermodynamic properties, and transport coefficients.
\end{itemize}

Using the blending technique and the newly derived EC fluxes, we construct an entropy-stable DGSEM (ES-DGSEM) scheme for the hypersonic Euler equations. The ES-DGSEM scheme is first evaluated and compared with other high-order spectral element methods for the single-species compressible Euler equations initialized with the Leblanc test problem. Each approach is then employed to simulate a modified shock-density wave interaction problem with a high-pressure ratio. The shock-density wave interaction problem assesses the high-order scheme in handling shock turbulence interactions. Subsequently, the three-species Sod problem with simple chemistry is used to compare the positivity-preserving and shock-capturing abilities of available high-order methods for solving the multi-species Euler equations with reaction source terms. Next, the new EC fluxes' global and local entropy conservation is verified by solving a periodic smooth initial condition test case. This is followed by the calculation of reaction source terms and thermodynamic properties, which are validated by solving a hypersonic Sod problem with real chemistry. The effect of real versus ideal chemistry models is investigated by employing a hypersonic Sod problem. The last benchmark problem is the simulation of chemically reactive Navier-Stokes equations behind the hypersonic shock wave to validate the transport properties of the dissociating gas model. We conclude by summarizing the most important findings of our work.

\section{Methodology}
\label{method}

\subsection{Governing Equations}
\label{Gov_eq}
In the current study we develop high-order methods that solve the non-dimensional 1D compressible reactive Navier-Stokes equations governing hypersonic flows. The non-dimensionalization is performed using free-stream conditions denoted by the ``$\infty$" subscript. The transport equations are:

\begin{equation}
\frac{\partial \mathbf{U}}{\partial t}+\frac{\partial \mathbf{F} }{\partial x}=\mathbf{S},
\label{vec_gov}
\end{equation}
where $\mathbf{U}$ and $\mathbf{F}$  are the solution and flux vectors, respectively. The flux vector is defined as the summation of the advective and viscous fluxes $\mathbf{F}=\mathbf{F}^a+\mathbf{F}^v$. The solution, advective, viscous fluxes, and source terms are:
\begin{equation}
\mathbf{U}=\begin{pmatrix}\rho_1\\\vdots\\\rho_{N_s}\\ \rho u \\ \rho E
\end{pmatrix}^T, \quad \mathbf{F}^a=\begin{pmatrix}\rho_1 u\\\vdots\\\rho_{N_s} u\\ \rho u^2+p \\ u(\rho E+p)
\end{pmatrix}^T,
\quad \mathbf{F}^v=\begin{pmatrix}-\frac{\partial J_1}{\partial x} \\\vdots\\-\frac{\partial J_{N_s}}{\partial x}\\ \frac{\partial \sigma}{\partial x} \\ -\frac{\partial q}{\partial x}+\frac{\partial (u\sigma)}{\partial x}
\end{pmatrix}^T,
\quad \mathbf{S}=\begin{pmatrix}\dot{\omega}_1 \\\vdots\\\dot{\omega_{N_s}}\\ 0 \\ 0
\end{pmatrix}^T
\label{eq:vec}
\end{equation}

% \begin{equation}
% \frac{\partial \rho_i}{\partial t}+\frac{\partial \rho_i u}{\partial x}=-\frac{\partial J_i}{\partial x}+\Dot{\omega}_i\quad \textrm{for}\quad i=1,\cdots,N_s,
%     \label{eq:species}
% \end{equation}

% \begin{equation}
% \frac{\partial \rho u}{\partial t}+\frac{\partial  \left(\rho u^2+p\right)}{\partial x}=\frac{\partial \sigma}{\partial x},
%     \label{eq:momentum}
% \end{equation}
% and
% \begin{equation}
% \frac{\partial E}{\partial t}+\frac{\partial  \left(u\left(E+p\right)\right)}{\partial x}=-\frac{\partial q}{\partial x}+\frac{\partial \left(u\sigma\right)}{\partial x}.
%     \label{eq:energy}
% \end{equation}
In Eq.~\eqref{eq:vec}, $t$ and $x$ are temporal and spatial coordinates. The terms $\rho$, $u$, and $p$ denote density, velocity, and pressure, respectively. The species density is denoted by $\rho_i$ and the total density is calculated as $\rho=\sum_{i=1}^{N_s}\rho_i$. $J_i$ is the mass diffusion flux for species $i$. The term $\Dot{\omega}_i$ indicates the species source term due to the chemical reactions. Equation \eqref{eq:vec} represents the transport equation for the $N_s$ species forming the gas mixture. The viscous stress tensor, heat flux, and total energy are given by:
\begin{equation}
    \sigma = -\frac{4}{3}\frac{\mu}{Re_{\infty}}\frac{\partial u}{\partial x},
    \label{viscous1}
\end{equation}

\begin{equation}
    q = -\frac{\kappa}{Re_{\infty}Pr_{\infty}Ec_{\infty}}\frac{\partial T}{\partial x}+\sum_{i=1}^{N_s}h_i J_i,
    \label{viscous2}
\end{equation}

\begin{equation}
   \rho E = \frac{1}{Ec_{\infty}}\rho e +\frac{1}{2}\rho u^2,
    \label{viscous}
\end{equation}
where $\mu$ denotes the gas mixture's dynamic viscosity, $\kappa$ is the gas mixture's thermal conductivity, $h_i$ is the enthalpy of species $i$, and $T$ indicates the temperature of the gas mixture. The Reynolds $Re_{\infty}$, Prandtl $Pr_{\infty}$, and Eckert $Ec_{\infty}$ numbers are then defined as

\begin{equation}
Re_{\infty}=\frac{\rho^*_{\infty}U^*_{\infty}L^*_{f}}{\mu^*_{\infty}},\quad Pr_{\infty}=\frac{\mu^*_{\infty}C^*_{p,\infty}}{\kappa^*_{\infty}},\quad Ec_{\infty}=\frac{a^{*2}_{\infty}}{C^*_{p,\infty}(\gamma_{\infty}-1)T^*_{\infty}}.
\label{eq:nondimnumbers}
\end{equation}
In Eq.~\eqref{eq:nondimnumbers}, the superscript $*$ denotes the dimensional value of the quantity. The constant pressure-specific heat coefficient is presented by $C^*_{p,\infty}$. The term $a^*_{\infty}$ is the speed of sound in the free stream condition, and $\gamma_{\infty}$ is the ratio of the specific heats. The term $ L^*_f$ denotes the reference length scale.

The Navier-Stokes equations are completed by adding a physico-chemical model representing a high-temperature gas mixture's thermodynamic and transport properties along with the species transport source terms due to the chemical reaction terms. The physico-chemical variables are functions of temperature, density, internal energy, mass fraction, and mole fractions as described below:

\begin{equation}
    \begin{array}{l}
      T=T^*(e^*,\rho^*)/\left((\gamma_{\infty}-1)T^*_{\infty}\right)   \\
      p=p^*\left(T^*,\rho^*,Y_i\right)/\left(\rho^*_{\infty}a^{*2}_{\infty}\right), 
         \\
      \mu = \mu^*\left(T^*,p^*,Y_i\right)/\mu^*_{\infty},
         \\
         \kappa=\kappa^*\left(T^*,p^*,Y_i,e^*\right)/\kappa^*_{\infty},
         \\
         h_i=h_i^*\left(T^*,p^*\right)/a^{*2}_{\infty},
         \\
         J_i=J_i^*\left(\partial X_i/\partial x^*, T^*,p^*,Y_i\right)/\left(\rho^*_{\infty}a^*_{\infty}\right),
         \\
         \Dot{\omega}_i=M_i\Omega^*_i\left(T^*,\rho^*,p^*,Y_i\right)/\left(\rho^*_{\infty}a^*_{\infty}/L^*_f\right),
    \end{array}
   \label{eq:physiochem} 
\end{equation}
where $M_i$ is the molar mass of species $i$, $Y_i$ is species $i$ mass fraction, and $X_i$ is the mole fraction of species $i$.

\subsection{Physico-Chemical Model}
\label{physiochem}
The thermodynamic and transport properties and the chemical reaction source terms are calculated using the Mutation++ library  \cite{Scoggins2020}. In the Mutation++ library, a simple Rigid-Rotor Harmonic-Oscillator (RRHO) model is employed to determine the thermodynamic properties of the gas mixture. The transport coefficients such as viscosity, thermal conductivity, and mass diffusivity are determined by solving the Boltzmann equation using the Chapman-Enskog perturbative method \cite{Scoggins2020}. 

\subsubsection{Thermodynamic properties}
\label{thermo}
In this section we consider a thermodynamic model for the five-species air mixture model where ionization does not occur. The species consists of $N$, $O$, $NO$, $N_2$, and $O_2$. The internal energy $e^*$ of the gas mixture is computed as the summation of all the gas species' internal energies weighted by their mass fractions as

\begin{equation}
e^*=\sum_{i=1}^{N_s}Y_i e^*_i,\quad e^*_i=e^*_{i,0}+e^*_{i,\textrm{trans}}+e^*_{i,\textrm{rot}}+e^*_{i,\textrm{vib}},
    \label{internal_energy}
\end{equation}
where $e^*_{i,0}$ is the formation enthalpy of the $i^\textrm{th}$ species. The specific transnational energy of species $i$ is computed as 
\begin{equation}
    e^*_{i,trans}=\frac{3}{2}\frac{R^*}{M_i}T^*, \quad i=1,\cdots,N_s,
\end{equation}
where $R^*$ is the universal gas constant. The vibrational and rotational energies vanish for mono-atomic species, while for diatomic molecules can be computed as 

\begin{equation}
e^*_{i,vib} = \Bigg\{\begin{array}{lc}
      0, & i \notin \mathcal{A} \\
      \frac{R^*}{M_i}\frac{\theta^*_i}{\exp{\left(\frac{\theta^*_i}{T^*}\right)}-1}, & i \in \mathcal{A}
\end{array}
    \label{vib}
\end{equation}

\begin{equation}
e^*_{i,rot} = \Bigg\{\begin{array}{lc}
      0, & i \notin \mathcal{A} \\
      \frac{R^*}{M_i}, & i \in \mathcal{A}
\end{array}
    \label{vib}
\end{equation}
where $\mathcal{A}$ is the subset of diatomic species and $\theta^*_i$ is the vibrational characteristic temperature associated with the single vibrational mode. The ideal gas equation of state is employed to calculate the pressure of the gas mixture as

\begin{equation}
    p^*=\sum_{i=1}^{N_s}\rho^*_i\frac{R^*}{M_i}T^*=\rho^*\sum_{i=1}^{N_s}R^*\frac{Y_i}{M_i}T^*
\end{equation}
where the species mass fraction is defined as $Y_i=\rho^*_i/\rho^*$. Species mole fractions and the gas mixture molecular weight can also be computed as

\begin{equation}
\chi_i=Y_i \frac{M}{M_i},
    \label{mole_frac}
\end{equation}
\begin{equation}
M=\left(\sum_{i=1}^{N_s}\frac{Y_i}{M_i}\right)^{-1}.
    \label{mole_frac}
\end{equation}
The enthalpy of each species is determined as
\begin{equation}
    h^*_i=e^*_i+\frac{2}{3}e^*_{trans,i}.
\label{enthalpy}    
\end{equation}

\subsubsection{Transport Coefficients}
\label{transport}
The transport coefficients, such as the dynamic viscosity, thermal conductivity, and mass diffusivity, are calculated based on an algorithm presented in \cite{MAGIN2004424}. The dynamic viscosity of the gas mixture is determined as 

\begin{equation}
    \mu^*=\sum_{i=1}^{N_s}\eta_i\chi_i,
\end{equation}
where $\eta_i$ is computed by solving the following linear system of equations 

\begin{equation}
    \sum_{i=1}^{N_s}G^*_{ij}\eta_i=\chi_j, \quad j=1,\cdots,N_s,
\end{equation}
where $G^*_{ij}$ is the viscosity matrix reported as $G^\eta$ in Appendix C of \cite{MAGIN2004424}. 

The determination of the thermal conductivity is similar to the work of Marxen \emph{et al.}~\cite{marxen2013method} where 

\begin{equation}
    \kappa^*=\kappa^*_{trans}+\kappa^*_{int},
\end{equation}
\begin{equation}
    \kappa_{trans}^*=\sum_{i=1}^{N_s}\xi_i\chi_i,
\end{equation}
and
\begin{equation}
    \sum_{i=1}^{N_s}K^*_{ij}\xi_i=\chi_j, \quad j=1,\cdots,N_s,
\end{equation}
where $K^*_{ij}$ is the thermal conductivity matrix presented in Appendix C of \cite{MAGIN2004424} denoted as $G^{\lambda_h}$. The internal thermal conductivity $\kappa^*_{int}$ is defined as

\begin{equation}
    \kappa^*_{int}=\kappa^*_{rot}+\kappa^*_{vib}=\sum_{j\in\mathcal{A}}\frac{\rho^*_i(C^*_{j,rot}+C^*_{j,vib})}{\sum_{i=1}^{N_s}\frac{\chi_i}{\mathcal{D}_{ji}}},
    \label{intthercond}
\end{equation}
where $C^*_{j,vib}=de^*_{vib,i}/dT^*$ and $C^*_{j,rot}=de^*_{rot,i}/dT^*$ are the vibrational and rotational species specific heats per unit mass, respectively; $D_{ji}$ are the binary mass diffusion coefficients and are presented in Appendix A of \cite{MAGIN2004424}, denoted by $\mathfrak{D}$. If we neglect the mass diffusion due to temperature and pressure gradients, the diffusion driving force could be described as 

\begin{equation}
    d^*_{i}=\frac{\partial X_i}{\partial x^*}.
    \label{massdiff}
\end{equation}
The mass diffusion flux is then computed as 
\begin{equation}
    J^*_{i}=\rho^*_i \tau^*_{i},
    \label{massdiff}
\end{equation}
\begin{equation}
    \sum_{i=1}^{N_s}Q^*_{ij}=-d^*_{j},\quad j=1,\cdots,N_s,
    \label{massdiff}
\end{equation}
with Stefan-Maxwell matrix $Q^*$ given in Appendix C of \cite{MAGIN2004424}, denoted by $G^V$. The gas mixture mass diffusion vanishes, meaning that $\sum_{i=1}^{N_s}J^*_i=0$.

\subsubsection{Chemistry}
\label{chem}

The reaction mechanism can be cast into an elementary reaction set as $\Lambda$ for all the species. The reactions can be modeled by

\begin{equation}
\sum_{i=1}^{N_s} \mathcal{C}_{\mathcal{R},\xi}^{i}X_i=\sum_{i=1}^{N_s} \mathcal{C}_{\mathcal{P},\xi}^{i}X_i,\quad \xi\in \Lambda,
    \label{reaction}
\end{equation}
where $\mathcal{C}_{\mathcal{R},\xi}^{i}$ and $\mathcal{C}_{\mathcal{P},\xi}^{i}$ indicate the stoichiometric coefficients of the $\xi^\textrm{th}$ reactions for reactants and products, respectively. The species source terms are then computed using 

\begin{equation}
\Omega^*_i = \sum_{\xi=1}^{\Lambda}\left(\mathcal{C}^i_{\mathcal{P},\xi}-\mathcal{C}^i_{\mathcal{R},\xi}\right)\left[k^*_{f,\xi}\prod_{m=1}^{N_s}\left(\frac{\rho^*_i}{M_i}\right)^{\mathcal{C}_{\mathcal{R},\xi}^i}-k^*_{b,\xi}\prod_{m=1}^{N_s}\left(\frac{\rho^*_i}{M_i}\right)^{\mathcal{C}_{\mathcal{P},\xi}^i}\right],\quad i=1,\cdots,N_s.
    \label{reaction}
\end{equation}
The forward reaction rate $k^*_{f,\xi}$ is obtained using the Arrhenius law as 

\begin{equation}
k^*_{f,\xi}=C^*_\xi T^{*{\beta_\xi}}\exp{\left(\frac{\nu_{\xi}}{k^*_B T^*}\right)},
    \label{arrhenius}
\end{equation}
where $C^*_\xi$ is a constant and $\beta_\xi$ is an integer number dependant on the reaction and $\nu_\xi$ denotes the activation energy for the $\xi^\textrm{th}$ reaction \cite{park2001chemical,park1993review}. The backward reaction rate can be determined using the following relation

\begin{equation}
k^*_{f,\xi}=k^*_{b,\xi}K^*_{e,\xi},\quad \xi \in \Lambda,
    \label{kbkf}
\end{equation}
where the equilibrium constant $K^*_{e,\xi}(T^*)$ is defined with a reference pressure, $p^*_{eq}=1 \,Pa$, as

\begin{equation}
\ln K^*_{e,\xi}=-\frac{1}{R^* T^*}\sum_{\xi=1}^{\Lambda}\left(\mathcal{C}^i_{\mathcal{P},\xi}-\mathcal{C}^i_{\mathcal{R},\xi}\right)g^*_i(T^*,p^*_{eq})M_i.
    \label{kedef}
\end{equation}
The Gibbs free energy is determined with 

\begin{equation}
g^*_i(T^*,p^*_{eq})=g^*_{i,T}+g^*_{i,R}+g^*_{i,V}.
    \label{gibbs}
\end{equation}
The transnational Gibbs free energy is defined as 

\begin{equation}
g^*_{i,T}(T^*,p^*_{eq})=-\frac{R^*}{M_i}T^*\ln{\left(\frac{R^*T^*}{N_A p^*_{eq}}\left[\frac{2\pi M_i R^* T^*}{N_A^2 h_p^{*2}}\right]^{3/2}\right)},\quad i=1,\cdots,N_s,
    \label{gibbs}
\end{equation}
where $N_A$ is Avogadro's number, and $h_p^{*}$ is Planck's constant. The specific rotational Gibbs free energy and specific vibrational Gibbs free energy are also defined as

\begin{equation}
g^*_{i,R}(T^*)=\begin{cases}
0 & i \notin \mathcal{A} \\ 
\frac{R^*}{M_i} T^* \ln{\left(\frac{\Theta_{i,R}\sigma_i}{T^*}\right)} & i \in \mathcal{A}, 
\end{cases}
\label{rot}
\end{equation}
and

\begin{equation}
g^*_{i,V}(T^*)=\begin{cases}
0 & i \notin \mathcal{A} \\- 
\frac{R^*}{M_i} T^* \ln{\left(1-\exp{\left(\frac{\Theta_{i,V}}{T^*}\right)}\right)} & i \in \mathcal{A}. 
\end{cases}
\label{vib}
\end{equation}
In Eq.~\eqref{rot}, $\Theta_{i,R}$ is the rotational characteristic temperature and $\sigma_i$ indicates the steric factor for species $i$. All the spectroscopic constants are reported in \cite{gurvich1989thermodynamics}. In this study, all the test cases with real chemistry employ the five species five reactions mechanism of Park \cite{park1993review}.

\subsection{Discontinuous Galerkin Spectral Element Method}
\label{dgsem}
In this study, we compare three spectral element methods: The DGSEM, SD, and FR schemes, The goal is to distinguish a high-order solver capable of simulating high-pressure ratio shock waves specific to the hypersonic flow simulations. DGSEM is the main numerical method that we extend to  hypersonic flows. Here, we describe the numerical framework and the stabilizing technique used for the numerical simulation of high-pressure ratio shock waves. We consider the governing equation of the $I^\textrm{th}$ conservative variable in Eq. \eqref{vec_gov} as  
\begin{equation}
\frac{\partial U^I}{\partial t}+\frac{\partial F^I(\mathbf{U},\mathbf{W}_x)}{\partial x}=S^I, \quad x\in\Omega
    \label{scalar}
\end{equation}
where $U^I=U^I(x,t)$ is the $I^\textrm{th}$ conservative variable solution, $F^I(\mathbf{U},\mathbf{W}_x)$ is the flux and $S^I$ is the source term for the corresponding $I^\textrm{th}$ conservative variable. The vector of primitive variables is also defined as $\mathbf{W}=(\rho_1,\cdots,\rho_{N_s},u,T)^T$. We
divide the physical domain, $\Omega$,  into $N$ non-overlapping and geometrically conforming elements defined as $\Omega \approx \Omega_k=\{x|x_k < x< x_{k+1}\}$ for $k=1,\cdots,N$. The global solution is approximated as a direct sum of the local piecewise polynomial solution and expressed as
\begin{equation}
U^I(x,t)\approx U^{eD,I}(x,t)=\bigoplus_{k=1}^{N}U_k^{eD,I}(x,t),
\label{sol_app}
\end{equation}  
whereas the flux function is estimated by a union of continuous elemental functions as
\begin{equation}
F^I(x,t)\approx F^{e,I}(x,t) = \bigoplus_{k=1}^{N}F_k^{e,I}(x,t).
\label{flux_app}
\end{equation}
In Eqs.~\eqref{sol_app} and \eqref{flux_app}, the superscript $e$ refers to the approximate solution or flux inside the $k^\textrm{th}$ element. From now on, we omit the superscripts $e$ and $D$ for brevity. An affine mapping is used to map 1D elements onto a reference elements $[-1,1]$ with reference coordinate $\eta$ to perform the numerical integration over the element. The mapping function is $x(\eta)=0.5(1-\eta)x_k+0.5(1+\eta)x_{k+1}$. Equation \eqref{scalar} is recast into the reference element as follows 
\begin{equation}
\frac{\partial \Tilde{U}_k^I}{\partial t}+\frac{1}{|J|}\frac{\partial \Tilde{F}^{I}_k}{\partial \eta}=\Tilde{S}^I_k,
    \label{mapped}
\end{equation}
where $|J|$ is local transformation Jacobian and defined as $|J|=x_\eta$ and the variables $\Tilde{U}_k^I$ and $\Tilde{F}^{I}_k$ denotes the respective quantity in the mapped space. To recover the weak form of \eqref{mapped}, we multiply Eq.~\eqref{mapped} with a Lagrange polynomial test function constructed based on Lobatto Gauss Legendre (LGL) points. These Lagrange polynomials are defined in the interval $\eta \in [-1,1]$ as 
\begin{equation}
l_i(\eta)=\prod_{m=0,m\neq i}^{\mathcal{P}} \frac{\eta-\eta_{m}}{\eta_{i}-\eta_{m}}.
\label{lag}
\end{equation}
We then integrate the equation over the interval $[-1,1]$ to derive

\begin{equation}
\frac{\partial }{\partial t}\int_{-1}^1 \Tilde{U}_k^I(\eta,t) l_i(\eta) d\eta+\frac{1}{|J|}\int_{-1}^1\frac{\partial \Tilde{F}^{I}_k(\eta,t)}{\partial \eta} l_i(\eta) d\eta=\int_{-1}^1 \Tilde{S}^I_k(\eta,t)l_i(\eta)d\eta.
    \label{integarlform}
\end{equation}
We perform integration by parts for the flux derivative and integral term and rewrite the equation as
\begin{equation}
 \frac{\partial }{\partial t}\int_{-1}^1 \Tilde{U}_k^I(\eta,t) l_i(\eta) d\eta=-\frac{1}{|J|}\left(\Tilde{F}^{I}_k(\eta,t)l_i(\eta)\Big|_{-1}^1-\int_{-1}^1\Tilde{F}^{I}_k(\eta,t) \frac{\partial l_i(\eta)}{\partial \eta} d\eta\right)+\int_{-1}^1 \Tilde{S}^I_k(\eta,t)l_i(\eta)d\eta.
    \label{sbpform}
\end{equation}
The Gauss-Lobatto quadrature rule is applied to integrals in \eqref{sbpform} to derive the discrete form as
\begin{equation}
  \sum_{j=0}^\mathcal{P} \dot{\Tilde{U}}_{k,j}^I l_i(\eta_j)\omega_j=-\frac{1}{|J|}\left(\Tilde{F}^{*I}_{r,k}-\Tilde{F}^{*I}_{l,k}-\sum_{j=0}^\mathcal{P}\Tilde{F}^{I}_{j,k}\frac{d l_i(\eta_j)}{d \eta} \omega_j\right)+\sum_{j=0}^\mathcal{P}\Tilde{S}^{I}_{j,k}l_i(\eta_j)\omega_j,
    \label{discrete}
\end{equation}
where $\omega_j$ is the quadrature weight for the $j^\textrm{th}$ quadrature point.
The semi-discrete form of Eq.~\eqref{discrete} is written as
\begin{equation}
    \frac{d \vec{U}_k^I}{d t}=-\frac{1}{|J|}\left(\mathbf{M}^{-1}\mathbf{B}\Vec{F}^{*I}_k-\mathbf{M}^{-1}\mathbf{D}^T\mathbf{M}\Vec{F}^I_k\right)+\Vec{S}^I_k.
    \label{vec_form}
\end{equation}
The boundary matrix is also defined as 

\begin{equation}
\mathbf{B}=\textrm{diag}([-1,0,\cdots,0,1]).
    \label{boundar}
\end{equation}
The superscript $*$ indicates the interface numerical flux calculated for advective and viscous fluxes at the element's boundaries. 
The mass and differentiation matrices satisfy the following property 
\begin{equation}
\mathbf{M}\mathbf{D}+(\mathbf{M}\mathbf{D})^T=\mathbf{B},
    \label{boundar}
\end{equation}
which is referenced to as the summation by parts (SBP) property \cite{doi:10.1137/120890144}. Using the SBP property, we can derive the equivalent strong form of the DGSEM scheme from \eqref{vec_form} as 
\begin{equation}
    \frac{d \Vec{U}^I_k}{d t}=-\frac{1}{|J|}\left(\mathbf{M}^{-1}\mathbf{B}\left(\Vec{F}^{*I}_k-\Vec{F}^I_k\right)-\mathbf{D}\Vec{F}^I_k\right)+\Vec{S}^I_k.
    \label{strongf}
\end{equation}
The matrices and vectors in Eq.~\eqref{strongf} are expressed as

\begin{equation}
\mathbf{M}[i,j]=\omega_j \delta_{j,i},\quad \mathbf{D}[j,i]=\frac{d l_i(\eta_j)}{d \eta},\quad U^I_{k,j}=\Tilde{U}_k^{I}(\eta_j,t),\quad F^I_{k,j}=\Tilde{F}_k^{I}(\eta_j,t),\quad S^I_{k,j}=\Tilde{S}_k^{I}(\eta_j,t).
\label{matrices}
\end{equation}
In Eq.~\eqref{strongf}, $\Vec{F}^{*I}_k=\left(F^{*I}_{l,k},0,\cdots,0,F^{*I}_{r,k}\right)^T$ consists of left and right boundaries fluxes. We will also decompose the flux function vectors into advective and diffusive parts such that $\Vec{F}^{*I}_k=\Vec{F}^{*I,a}_k+\Vec{F}^{*I,v}_k$ and $\Vec{F}^{I}_k=\Vec{F}^{I,a}_k+\Vec{F}^{I,v}_k$. 

\subsubsection{Advective fluxes}
\label{advect}
The advective interface fluxes are determined by constructing a Riemann problem from the two different states of the solution vectors at the left and right-hand sides of the interface. The Riemann problem is then solved using an approximate Riemann solver such as HLLC \cite{Toro:book} to determine the common interface flux. Therefore, the interface fluxes are calculated as 

\begin{equation}
F^{*I,a}_{l,k}=F^{HLLC}(\mathbf{U}_{k-1,\mathcal{P}},\mathbf{U}_{k,0}),\quad F^{*I,a}_{r,k}=F^{HLLC}(\mathbf{U}_{k,\mathcal{P}},\mathbf{U}_{k+1,0}).
    \label{interfacerimann}
\end{equation}
For the advective volume fluxes, Fisher \emph{et al.}~\cite{fisher2013high} and Carpenter \emph{et al.}~\cite{carpenter2014entropy} proved that the derivative operator with the diagonal SBP property can be reformulated into a sub-cell based finite volume scheme as 

\begin{equation}
\sum_{n=0}^\mathcal{P}D_{in}F^{I,a}_{k,n}=\frac{\Bar{F}^{I,a}_{k,i+1}-\Bar{F}^{I,a}_{k,i}}{\omega_i},\quad i=0,\cdots,\mathcal{P},\quad\textrm{and}\quad k=1,\cdots,N.
    \label{differencing}
\end{equation}
The flux differencing from Eq. \eqref{differencing} proves the conservation of the derivative operator. Fisher \emph{et al.}~\cite{fisher2013high} and Carpenter \emph{et al.}~\cite{carpenter2014entropy} showed that the volume flux differencing relation \eqref{differencing} can be written in terms of two-point entropy conserving flux functions $F^{**I}_{EC}=F^{**I}_{EC}(\mathbf{U}_{k,i},\mathbf{U}_{k,j})$ as

\begin{equation}
\frac{\Bar{F}^{I,a}_{k,i+1}-\Bar{F}^{I,a}_{k,i}}{\omega_i}=2\sum_{n=0}^\mathcal{P}D_{in}F^{**I}_{k,EC}(\mathbf{U}_{k,i},\mathbf{U}_{k,n}).
    \label{fluxdiff}
\end{equation}
Fisher \emph{et al.}~\cite{fisher2013high} and Carpenter \emph{et al.}~\cite{carpenter2014entropy} then showed that the two-point entropy conservative flux results in  a high-order accurate discretization. Gassner \emph{et al.}~\cite{gassner2016split} later concluded that the high order accuracy resulted from the symmetry and consistency of the two-point flux function and not from the entropy conservation. Therefore, they rewrote Eq.~\eqref{fluxdiff} such that

\begin{equation}
\frac{\Bar{F}^{I,a}_{k,i+1}-\Bar{F}^{I,a}_{k,i}}{\omega_i}=2\sum_{n=0}^\mathcal{P}D_{in}F^{**I}_{k}(\mathbf{U}_{k,i},\mathbf{U}_{k,n}),
    \label{fluxdiffcons}
\end{equation}
where the two-point flux function $F^{**I}_{k}$ satisfies consistency and symmetry conditions as 
\begin{equation}
    F^{**I}_k(\mathbf{U}_{k,i},\mathbf{U}_{k,i})=F^I_k(\mathbf{U}_{k,i})\quad\textrm{and}\quad F^{**I}_k(\mathbf{U}_{k,i},\mathbf{U}_{k,j})=F^{**I}_k(\mathbf{U}_{k,j},\mathbf{U}_{k,i}).
    \label{conssymm}
\end{equation}
Two-point entropy conservative fluxes can be determined using the semi-discrete setting of Tadmor \cite{tadmor1987numerical,tadmor2003entropy}. We derive new entropy conservative fluxes \cite{ranocha2018comparison} for the hypersonic equations in Section \ref{entconv}. Here, we can also employ the split form fluxes presented in the literature such as standard DG \cite{kopriva2009implementing}, Morinishi \cite{morinishi2010skew}, Ducros \emph{et al.}~\cite{ducros2000high}, Kennedy and Gruber \cite{kennedy2008reduced}, and Pirozzoli \cite{pirozzoli2011numerical}. Gassner \emph{et al.}~\cite{gassner2016split} presents the two-point flux formulations of all the split forms. This strategy avoids the tedious calculations of the entropy conservative fluxes for the hypersonic formulations. It also provides a flux differencing framework that can be exploited for constructing a conservative hybrid discretization for shock capturing.  

Using Eq.~\eqref{differencing}, the semi-discrete form of DGSEM \eqref{strong} is expressed as 

\begin{equation}
    \frac{d \vec{U}^I_k}{d t}=-\frac{1}{|J|}\left(\mathbf{M}^{-1}\mathbf{B}\left(\Vec{F}^{*I,a}_k-\Vec{F}^{I,a}_k\right)-\mathbf{M}^{-1}\Delta \Vec{\Bar{F}}^{I,a}_k\right)-\frac{1}{|J|}\left(\mathbf{M}^{-1}\mathbf{B}\left(\Vec{F}^{*I,v}_k-\Vec{F}^{I,v}_k\right)-\mathbf{D}\Vec{F}^{I,v}_k\right)+\Vec{S}^I_k,
    \label{strong}
\end{equation}
with 
\begin{equation}
\Delta=\begin{pmatrix}-1&1&0&\cdots&\cdots&0\\0&-1&1&0&\cdots&0\\\vdots&\ddots&\ddots&\ddots&\ddots&\vdots\\0&\cdots&0&-1&1&0\\0&\cdots&\cdots&0&-1&1
\end{pmatrix}\in \mathbb{R}^{(\mathcal{P}+1)\times(\mathcal{P}+2)},
    \label{finitevolumeder}
\end{equation}
and the high-order flux vectors $\Vec{\Bar{F}}_k^{I,a}$ are defined as
\begin{equation}
\Bar{F}^{I,a}_{k,0}=F^{I,a}_{k,0},
\quad
\Bar{F}^{I,a}_{k,j}=\sum_{m=j}^{\mathcal{P}}\sum_{n=0}^{j-1}2H_{nm}F^{**,I}_{k,(n,m)}\quad j=1,\cdots,\mathcal{P},
\quad
\Bar{F}^{I,a}_{k,\mathcal{P}+1}=F^{I,a}_{k,\mathcal{P}},
\label{DGvec}
\end{equation}
where $\mathbf{H}:=\mathbf{M}\mathbf{D}$. We can now blend the high-order flux $\Vec{\Bar{F}}^{I,a}_k$ with a first-order finite volume flux constructed on the interfaces of sub-cells. The sizes of those finite volume sub-cells are given by the quadrature weights of each LGL point \cite{hennemann2021provably}. In the sub-cells, the solution values on the LGL points are taken to be the sub-cell solution averages \cite{hennemann2021provably}. To stabilize the high-order scheme, we blend the high-order flux function with the low-order finite volume one. Using a combination of the high and low order fluxes, we rewrite Eq.~\eqref{strong} as

\begin{equation}
    \frac{d \vec{U}^I_k}{d t}=-\frac{1}{|J|}\left(\mathbf{M}^{-1}\mathbf{B}\left(\Vec{F}^{*I,a}_k-\Vec{F}^{I,a}_k\right)-\mathbf{M}^{-1}\Delta \left((1-\alpha)\Vec{\Bar{F}}^{I,a}_k+\alpha \Vec{F}^{I,FV}_k\right)\right)-\frac{1}{|J|}\left(\mathbf{M}^{-1}\mathbf{B}\left(\Vec{F}^{*I,v}_k-\Vec{F}^{I,v}_k\right)-\mathbf{D}\Vec{F}^{I,v}_k\right)+\Vec{S}^I_k
    \label{strongb}
\end{equation}
with $\alpha$ as the blending coefficient defined as $\alpha \in [0,\alpha_{max}]$. The blending coefficient $\alpha$ is constant over each element. The limit $\alpha_{max}$ is a user-defined value that satisfies $\alpha_{max}\le 1$. The value of $\alpha$ is determined based on the decay rate of the energy stored at high modes on each element \cite{hennemann2021provably}. The finite volume flux vector is expressed as

\begin{equation}
\Vec{F}_k^{I,FV}=\begin{pmatrix}0\\\Tilde{F}_k^{*I,a}\left(\mathbf{U}_{k,0},\mathbf{U}_{k,1}\right)\\\vdots\\ \Tilde{F}_k^{*I,a}\left(\mathbf{U}_{k,\mathcal{P}-1},\mathbf{U}_{k,\mathcal{P}}\right)\\ 0
\end{pmatrix}\in \mathbb{R}^{\mathcal{P}+2}
    \label{finitevolumeflux}
\end{equation}
where the $*$ superscript denotes the Riemann flux calculated from the two solution states at an interface. When $\alpha=1$, the discretization reduces to a finite volume scheme, and when $\alpha=0$, the discretization reverts to the high-order DGSEM scheme. The blending technique is identical to the work of Hennemann \emph{et al.}~\cite{hennemann2021provably}.

\subsubsection{Viscous fluxes}
\label{visfl}
In Eq.~\eqref{strongb}, the viscous flux vectors are expressed as 

\begin{equation}
F^{I,v}_{k,j}=F^{I,v}_{k,j}(\mathbf{U}_{k,j},\mathbf{Q}_{k,j}),
    \label{visfluxcalc}
\end{equation}
where $\mathbf{Q}_{k,j}=\mathbf{W}_{x,k,j}$ is the first order derivative of the primitive variable. We first define the equation $Q^I=W^I_x$ and map it onto the reference element. We then formulate the weak form similar to Eqs.~\eqref{integarlform}-\eqref{discrete} as

\begin{equation}
\sum_{j=0}^\mathcal{P} \Tilde{Q}_{k,j}^I l_i(\eta_j)\omega_j=-\frac{1}{|J|}\left(\Tilde{W}^{*I}_{r,k}-\Tilde{W}^{*I}_{l,k}-\sum_{j=0}^\mathcal{P}\Tilde{W}^{I}_{j,k}\frac{d l_i(\eta_j)}{d \eta} \omega_j\right).
    \label{auxil}
\end{equation}
The equivalent strong form of Eq.~\eqref{auxil} reads as
\begin{equation}
\Vec{Q}^I_k=-\frac{1}{|J|}\left(\mathbf{M}^{-1}\mathbf{B}\left(\Vec{W}^{*I}_k-\Vec{W}^I_k\right)-\mathbf{D}\Vec{W}^I_k\right),
    \label{auxstrongf}
\end{equation}
with $\Vec{W}^{*I}_k=\left(W^{*I}_{l,k},0,\cdots,0,W^{*I}_{r,k}\right)^T$ where the common interface values are determined using an arithmetic mean such that $W^{*I}_{l,k}=\frac{1}{2}\times(W^{I}_{k-1,\mathcal{P}}+W^{I}_{k,0})$ and $W^{*I}_{r,k}=\frac{1}{2}\times(W^{I}_{k,\mathcal{P}}+W^{I}_{k+1,0})$. After calculating the auxiliary variable $\Vec{Q}^I_k$, we employ the relations \eqref{eq:vec}-\eqref{viscous2} to determine the viscous fluxes at the LGL points. For the viscous interface fluxes, we enforce continuity by averaging the fluxes across the interface.

\section{Entropy conservative flux}
\label{entconv}
In this section, we derive two-point entropy conservative fluxes for formulation \eqref{internal_energy} of the hypersonic flows' internal energy. First, we write the internal energy equation based on Eq.~\eqref{internal_energy} as

\begin{equation}
\rho e = \sum_{k=1}^{N_s} \rho_k e_k, \quad e_k=e^0_k+a_k r_k T+\alpha_k r_k \frac{\theta_k}{e^{\frac{\theta_k}{T}}-1},
    \label{internalenergy}
\end{equation}
where $r_k=R/M_k$ and 
\begin{equation}
a_k=\Bigg\{\begin{array}{lc}
         \frac{3}{2}& \textrm{if}\quad k\notin \mathcal{A} \\
        \frac{5}{2} & \textrm{if}\quad k\in \mathcal{A} 
    \end{array},\quad \alpha_k=\Bigg\{\begin{array}{lc}
         0& \textrm{if}\quad k\notin \mathcal{A} \\
        1 & \textrm{if}\quad k\in \mathcal{A}
    \end{array}.
    \label{akalk}
\end{equation}
The constant volume specific heat is determined by taking the derivative of Eq.~\eqref{internalenergy} with respect to temperature as
\begin{equation}
c_{vk}=\frac{d e_k}{d T}=a_k r_k+\alpha_k r_k \theta_k \frac{-\frac{\theta_k}{T^2}e^{\frac{\theta_k}{T}}}{\left(e^{\frac{\theta_k}{T}}-1\right)^2}.
    \label{spce}
\end{equation}
According to Gouasmi \emph{et al.}~\cite{gouasmi2020formulation}, the numerical entropy for multi-species gas mixtures can be defined as 
\begin{equation}
\rho s = \sum_{k=1}^{N_s}\rho_k s_k,\quad s_k=\int \frac{c_{vk}(\tau)}{\tau}d\tau-r_k\ln{(\rho_k)}.
    \label{enteq}
\end{equation}
In Eq.~\eqref{enteq}, we substitute $c_{vk}$ with its expression from Eq.~\eqref{spce} and determine the species entropy as
\begin{equation}
s_k=a_k r_k \ln{(T)}+\frac{\alpha_k r_k \theta_k }{T}+\alpha_k r_k \theta_k\frac{\frac{1}{T}}{e^{\frac{\theta_k}{T}}-1}-\alpha_k r_k \ln{(e^{\frac{\theta_k}{T}}-1)}-r_k \ln{(\rho_k)}.
    \label{entcal}
\end{equation}
The entropy variables are the derivative of the entropy function \eqref{enteq} with respect to the conservative variables ($\omega_i=
\frac{\partial (\rho s)}{\partial \mathbf{U}_i}$). They can also be expressed based on the Gibbs function \cite{gouasmi2020formulation} as 
\begin{equation}
\mathbf{\omega}=\frac{1}{T}\left(g_1-\frac{1}{2}u^2,\cdots,g_{N_s}-\frac{1}{2}u^2,u,-1\right)^T.
    \label{entvar}
\end{equation}
The expressions for the entropy variables simplify into 
\begin{equation}
\omega_{1,k}=\frac{e_k}{T}-s_k-\frac{u^2}{2T},\quad \omega_2=\frac{u}{T},\quad \omega_3=\frac{-1}{T},
    \label{entvarr}
\end{equation}
where $k=0,\cdots,N_s$. We can also derive the entropy potential functions $\phi$ and $\psi$ as
\begin{equation}
    \phi=\sum_{k=1}^{N_s}r_k \rho_k,\quad \psi=\sum_{k=1}^{N_s}\rho_k r_k u.
    \label{potenfunc}
\end{equation}
According to Tadmor relation \cite{tadmor1987numerical,tadmor2003entropy}, the entropy conservative fluxes satisfy the following condition

\begin{equation}
[\![\omega]\!].\mathbf{F}^{EC}-[\![\psi]\!]=0,
    \label{tadmorcond}
\end{equation}
where the $\mathbf{F}^{EC}=(f^{EC}_{1,1},\cdots,f^{EC}_{1,N_s}, f^{EC}_2,f^{EC}_3)^T$ and $\omega$ is the entropy variables vector. The jump and average operators are expressed as
\begin{equation}
[\![a]\!]=a^+-a^-,\quad \{\!\{a\}\!\}=\frac{1}{2}(a^++a^-).
    \label{jumpave}
\end{equation}
We now need to define the jump conditions for entropy variables and for the potential function $\psi$ \eqref{tadmorcond}. First, we define the auxiliary variables as $z_{1,k}=\rho_k$, $z_2=u$, and $z_3=\frac{1}{T}$ and rewrite the entropy variables as 

\begin{equation}
    \omega_{1,k}=e^0_k z_3+a_k r_k +a_k r_k \ln{(z_3)}-\alpha_k r_k \theta_k z_3+\alpha_k r_k \ln{(e^{\theta_k z_3}-1)}+r_k \ln{(z_{1,k})}-\frac{1}{2} z_3 z_2^2,
    \label{entvarz}
\end{equation}
\begin{equation}
\omega_2=z_2 z_3,
    \label{omeg2}
\end{equation}
\begin{equation}
\omega_3=-z_3.
    \label{omeg3}
\end{equation}
We then determine the jumps of entropy variables as follows
\begin{equation}
    [\![\omega_{1,k}]\!]=e^0_k [\![z_3]\!]+a_k r_k \frac{[\![z_3]\!]}{\{\!\{z_3\}\!\}^{ln}}-\alpha_k r_k \theta_k [\![z_3]\!]+ \alpha_k r_k \frac{[\![z_3]\!]}{\{\!\{z_3\}\!\}^{ln,exp,k}}+r_k \frac{[\![z_{1,k}]\!]}{\{\!\{z_{1,k}\}\!\}^{ln}}-\frac{1}{2} \{\!\{z_2^2\}\!\} [\![z_{3}]\!]-\{\!\{z_2\}\!\}\{\!\{z_3\}\!\}[\![z_{2}]\!],
    \label{entvarz}
\end{equation}
\begin{equation}
[\![\omega_{2}]\!]=\{\!\{z_2\}\!\}[\![z_{3}]\!]+\{\!\{z_3\}\!\}[\![z_{2}]\!],
    \label{omeg2entjump}
\end{equation}
\begin{equation}
[\![\omega_{3}]\!]=-[\![z_{3}]\!],
    \label{omeg3entjump}
\end{equation}
where
\begin{equation}
\left[\!\left[\ln\left(a\right)\right]\!\right]=\frac{[\![a]\!]}{{\{\!\{a\}\!\}^{ln}}}.
    \label{aveages2}
\end{equation}
The logarithmic average operator is defined as
\begin{equation}
 \{\!\{a\}\!\}^{ln}=\frac{a^+-a^-}{\ln{(a^+)}-\ln{(a^-)}}.
    \label{jumpave1}
\end{equation}
Equation \eqref{jumpave1} must be implemented such that when $a^+ \rightarrow a^-$, the logarithmic average approaches a finite value. Ismail and Roe derive an accurate approximation \cite{ismail2009affordable} to determine the average value of Eq. \eqref{jumpave1} when the right and left states are close to each other. We need to derive the jump of the following logarithmic term 
\begin{equation}
\left[\!\left[\ln\left(e^{\theta_k a}-1\right)\right]\!\right]=\frac{[\![a]\!]}{{\{\!\{a\}\!\}^{ln,exp,k}}}.
    \label{aveages}
\end{equation}
We begin with defining an auxiliary variable $\chi$ as
\begin{equation}
\left[\!\left[\ln{\left(e^{\theta_k a}-1\right)}\right]\!\right]=\left[\!\left[\ln{\left(\chi\right)}\right]\!\right]=\frac{\left[\!\left[\chi\right]\!\right]}{\{\!\{\chi\}\!\}^{ln}}.
    \label{jumplen}
\end{equation}
We now formulate the jump of the auxiliary variable $\chi$ using the following relation
\begin{equation}
    \left[\!\left[\chi\right]\!\right]=\left[\!\left[e^{\theta_k a}\right]\!\right]=\frac{\left[\!\left[a\right]\!\right]}{\{\!\{a\}\!\}^{exp,k}}
\end{equation}
where
\begin{equation}
\{\!\{a\}\!\}^{exp,k}=\frac{a^+-a^-}{e^{\theta_k a^+}-e^{\theta_k a^-}}.
    \label{exp}
\end{equation}
As a result, the new logarithmic average is derived as
\begin{equation}
    {\{\!\{a\}\!\}^{ln,exp,k}}=\{\!\{e^{\theta_k a}-1\}\!\}^{ln}\{\!\{a\}\!\}^{exp,k}.
    \label{ave}
\end{equation}
We now explain how we can stably approximate the exponential mean average $\{\!\{a\}\!\}^{exp,k}$. We write the right and left states \cite{winters2020entropy} as

\begin{equation}
a^+ = \frac{a^++a^-}{2}+\frac{a^+-a^-}{2},\quad
a^- = \frac{a^++a^-}{2}-\frac{a^+-a^-}{2}
    \label{leftright}
\end{equation}
where an auxiliary variable $\xi$ can be defined as 
\begin{equation}
\xi = \frac{a^+-a^-}{a^++a^-}
    \label{aux}
\end{equation}
and rewrite Eq.~\eqref{leftright} as 
\begin{equation}
a^+ = \{\!\{a\}\!\}(1+\xi),\quad
a^- = \{\!\{a\}\!\}(1-\xi).
    \label{leftrightee}
\end{equation}
We substitute expressions of Eq.~\eqref{leftrightee} into Eq.~\eqref{exp} and derive
\begin{equation}
\{\!\{a\}\!\}^{exp,k}=\frac{\{\!\{a\}\!\}}{e^{\theta_k \{\!\{a\}\!\}}}\frac{\xi}{\sinh{\left(\theta_k \xi\right)}}.
    \label{expsim}
\end{equation}
We write the Taylor expansion of the hyperbolic sinh and simplify the expression to get the following approximation
\begin{equation}
\{\!\{a\}\!\}^{exp,k}=\frac{1}{\theta_k e^{\theta_k \{\!\{a\}\!\}}}\frac{1}{1+\frac{(\theta_k \{\!\{a\}\!\}\xi)^2}{3!}+\frac{(\theta_k \{\!\{a\}\!\}\xi)^4}{5!}+\frac{(\theta_k \{\!\{a\}\!\}\xi)^6}{7!}+\mathcal{O}((\theta_k\{\!\{a\}\!\}\xi)^9)}.
    \label{expend}
\end{equation}
We determine the jump in the potential function $\psi$ as
\begin{equation}
[\![\psi]\!]=\sum_{k=1}^{N_s}r_k\left(\{\!\{z_{1,k}\}\!\}[\![z_{2}]\!]+\{\!\{z_{2}\}\!\}[\![z_{1,k}]\!]\right).
    \label{potjump}
\end{equation}

Using Eq.~\eqref{tadmorcond}, we can determine the entropy conservative fluxes for the hypersonic flow as 
\begin{equation}
f_{1,k}=\{\!\{u\}\!\}\{\!\{\rho_k\}\!\}^{ln},
    \label{ECfluxf1}
\end{equation}

\begin{equation}
f_{2}=\frac{1}{\{\!\{\frac{1}{T}\}\!\}}\sum_{k=1}^{N_s}r_k\{\!\{\rho_k\}\!\}+\{\!\{u\}\!\}\sum_{k=1}^{N_s}f_{1,k},
    \label{ECfluxf2}
\end{equation}

\begin{equation}
f_{3}=\sum_{k=1}^{N_s}\left(e^0_k-\alpha_k r_k \theta_k+ \frac{a_k r_k}{\{\!\{\frac{1}{T}\}\!\}^{ln}}+\frac{\alpha_k r_k}{\{\!\{\frac{1}{T}\}\!\}^{ln,exp,k}} -\frac{1}{2}\{\!\{u^2\}\!\}\right)f_{1,k}+\{\!\{u\}\!\}f_2.
    \label{ECfluxf3}
\end{equation}
The entropy conservative fluxes are kinetic energy preserving besides being entropy conservative \cite{ranocha2018comparison}.

% We define a variable as $\xi=a^-/a^+$ and rewrite Eq.~\eqref{exp} as 

% \begin{equation}
% \{\!\{a\}\!\}^{exp,k}=\frac{a^+(1-\xi)}{e^{\theta_k a^+}\left(1-e^{-\theta_k(1-\xi)a^+}\right)}= \frac{a^+x}{e^{\theta_k a^+}\left(1-e^{-\theta_k x a^+}\right)}
%     \label{stableexpavearge}
% \end{equation}
% when $x=1-\xi$ approaches zero we can approximate the exponential term using the Taylor expansion calculated at zero as

% \begin{equation}
%   e^{-\theta_k x a^+}=1-\frac{\theta_k a^+ x}{1!}+\frac{\left(\theta_k a^+ x\right)^2}{2!}-\frac{\left(\theta_k a^+ x\right)^3}{3!}+\frac{\left(\theta_k a^+ x\right)^4}{4!}+\mathcal{O}(x^5).
%   \label{expappx}
% \end{equation}
% We substitute Eq. \eqref{expappx} into Eq. \eqref{stableexpavearge} and simplify to get
% \begin{equation}
% \{\!\{a\}\!\}^{exp,k}=\frac{1}{\theta_k e^{\theta_k a^+}}\frac{1}{1-\frac{\theta_k a^+ x}{2!}+\frac{\left(\theta_k a^+ x\right)^2}{3!}-\frac{\left(\theta_k a^+ x\right)^3}{4!}}.
%     \label{stableexp}
% \end{equation}
% Finally the last average is expressed as 
% \begin{equation}
%  \{\!\{a\}\!\}^{ln,k}= \{\!\{a\}\!\}^{exp,k}\{\!\{e^{\theta_k z_3}-1\}\!\}^{ln}.
%     \label{avelast}
% \end{equation}

\section{Results and discussion}
\label{resultdiscussion}
In this section, we present six benchmark problems to evaluate the performance of high-order methods simulating strong shock waves in flows near-vacuum conditions, strong shock-turbulence interactions, and strong shock waves inside chemically reactive flows. The validity of the entropy conservation of the hypersonic EC fluxes is shown. The coupling of the Mutation++ library with the ES-DGSEM scheme is verified by solving a hypersonic Sod problem with the reaction source terms, and a hypersonic viscous flow field behind a shock wave. For all the test cases, the Runge-Kutta 45 method of Carpenter and Kennedy \cite{CK94a} is employed for the time discretization.

\subsection{Benchmark 1: Leblanc Sod problem}

First, we considered the Leblanc Sod problem \cite{zhang2010positivity,fu2019very} to evaluate the positivity preservation and stabilization of common high-order spectral element methods. The initial conditions are defined as 

\begin{equation}
\left(\rho, u , p\right)=\begin{cases}
\left(2,0,10^9\right) & x \le 0.0 \\ 
\left(0.001,0,1\right) & x > 0.0, 

\end{cases}
\label{leblanc}
\end{equation}
where the pressure ratio is $10^9$ and the density ratio is $10^5$. The spatial domain is defined as $x\in[-10, 10]$ and the final time of the simulation is $t_{final}=0.0001$. The extreme high-pressure ratio in this test case is specific to the hypersonic flows where strong shock waves occur in near-vacuum conditions. For this test case, we solve the 1D Euler system of equations for a single species gas mixture with a constant $\gamma=1.4$.

\begin{figure}[t!]
  \begin{center}
    \begin{tabular}{cc}
    \includegraphics[width=0.45\textwidth,height=0.35\textwidth]{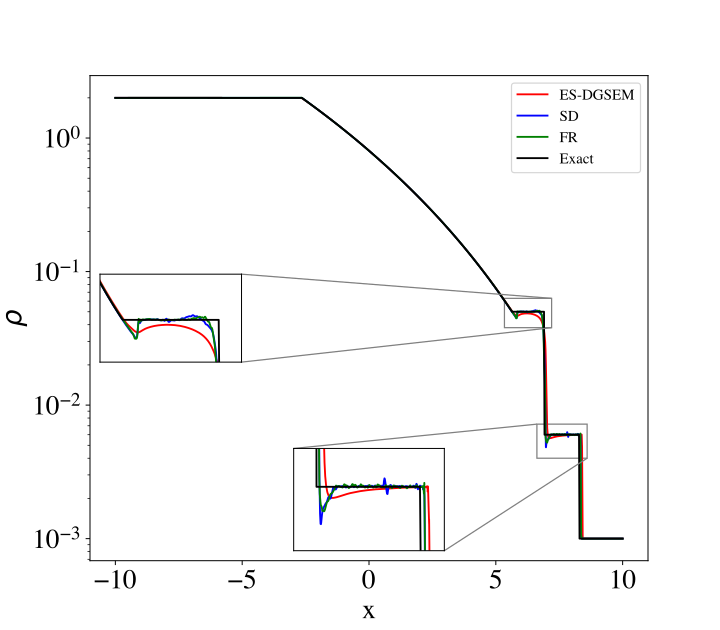}
 &
 \includegraphics[width=0.45\textwidth,height=0.35\textwidth]{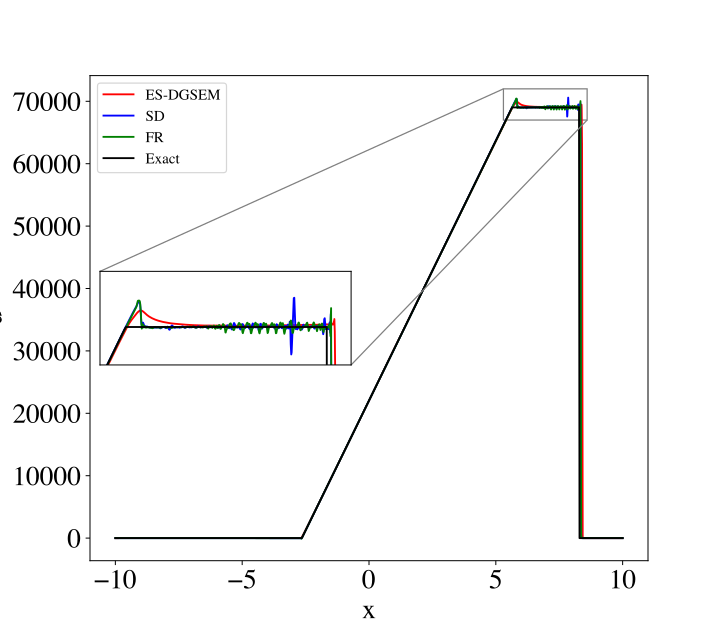}
    
  \\
  (a) Density& (b) Velocity
  \\
  \multicolumn{2}{c}{
   \includegraphics[width=0.45\textwidth,height=0.35\textwidth]{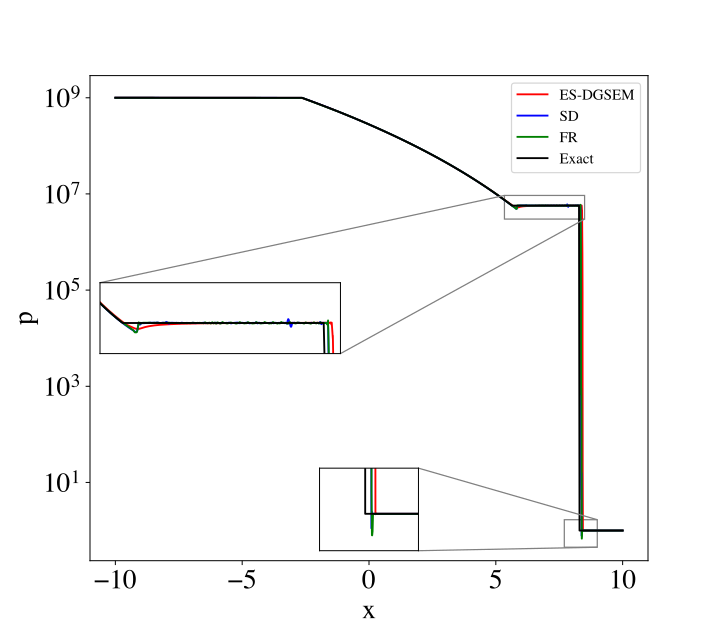}
   
  }
 \\
\multicolumn{2}{c}{(c) Pressure}

\end{tabular} 

\caption{Benchmark 1: Simulations of the Leblanc Sod problem corresponding to extreme conditions; the pressure ratio between the driver and driven sections is $10^9$ and the density ratio is $10^5$. Results are shown for ES-DGSEM (red solid line), SD (blue solid line), FR (green solid line) schemes against the exact solution (black solid line). The number of elements for all the cases is $1,000$ and the polynomial order is $\mathcal{P}=3$. (a) Density, (b) Velocity, and (c) Pressure at $t=0.0001$.}
    
    \label{fig:leblanc}
  \end{center}
  
\end{figure}

Figures~\ref{fig:leblanc}(a)-(c) present the density, velocity, and pressure profiles of the Leblanc Sod problem at $t=0.0001$. All the test cases employ $1000$ spectral elements with polynomial order $\mathcal{P}=3$. The time step size of all the simulations is fixed at $\Delta t=2.0893\times 10^{-8}$. The maximum allowable value for the blending coefficient is $\alpha_{max}=0.5$ for the ES-DGSEM scheme. The EC flux of Chandrashekar \cite{chandrashekar2013kinetic} is employed for the single species gas model in the ES-DGSEM scheme. According to Figure~\ref{fig:leblanc}(a), the ES-DGSEM scheme smears the contact discontinuity more than the FR and SD schemes since it detects shocks over an element. However, for the FR and SD schemes, the shock can be located between two solution points; therefore, the discontinuities are captured sharper than in the ES-DGSEM scheme. Figure~\ref{fig:leblanc}(b) reveals that the FR and SD approaches generate a higher amplitude of numerical oscillations close to the shock fronts compared to the ES-DGSEM scheme. Among the three schemes, the SD solver generates larger oscillations and undershoots at contact wave and shock front in density and pressure profiles. The discontinuous flux function in the SD scheme is constructed using $\mathcal{P}+1$ polynomial order, compared to $\mathcal{P}$ order polynomial in the FR and ES-DGSEM schemes. A higher polynomial degree can induce a higher amplitude of oscillations when evaluated at the boundaries of elements.

 A robust shock-capturing technique is required to resolve strong shock waves where stabilization can be achieved by filtering the solution down to mean modes \cite{dzanic2022positivity}. Adaptive modal filtering can lower the order of accuracy to one by zeroing out all the other modes except the mean mode. Therefore, for stability, the numerical scheme must ensure that the mean solution always stays within the physical bounds given a bounded solution value on all the solution points at the previous time step. All three high-order schemes maintain the physical bounds of the mean solution. According to our experience, employing artificial viscosity cannot stabilize the high-order spectral method for high-pressure ratios shock waves. The ES-DGSEM scheme performs robustly and generates the most negligible numerical oscillations in capturing large pressure ratio shock waves. At the same time, it slightly smears the sharp solution jumps compared to the FR and SD schemes. 
%%%%%%%%%%

\subsection{Benchmark 2: Shu-Osher shock-density wave interaction problem}

We consider the shock-density wave interaction problem to evaluate how the high-order scheme captures high-gradient continuous regions of the solution in the presence of shock waves \cite{shu1988efficient}. We have modified the original test problem to initialize a $M_s=10$ shock wave in the domain to interact with smooth density waves. Therefore, the initial conditions are defined as

\begin{equation}
\left(\rho, u , p\right)=\begin{cases}
\left(3.857143,9.76153,116.5\right) & x \le -4.0 \\ 
\left(1+0.2\sin{(5x)},0,1\right) & x > -4.0, 

\end{cases}
\label{shuosherm}
\end{equation}
where the physical domain is defined as $x \in [-5,5]$, the air is modeled as a calorically perfect single species gas with a constant $\gamma=1.4$. The initial conditions evolve in time until $t=0.7$, when the shock front reaches $x=4.27$, and high-frequency density waves form behind the shock. A high-resolution case with $N=3000$ elements and $\mathcal{P}=3$ is simulated using the DGSEM Euler solver to be considered an exact solution for comparison. All the simulation cases are performed with $N=170$ elements and $\mathcal{P}=3$. A constant time step size $\Delta t = 9.9964\times 10^{-5}$ is employed for all simulations to evaluate the accuracy of the results for a fixed number of time steps.

\begin{figure}[!htbp]
  \begin{center}
    \begin{tabular}{cc}
     \includegraphics[width=0.48\textwidth,height=0.38\textwidth]{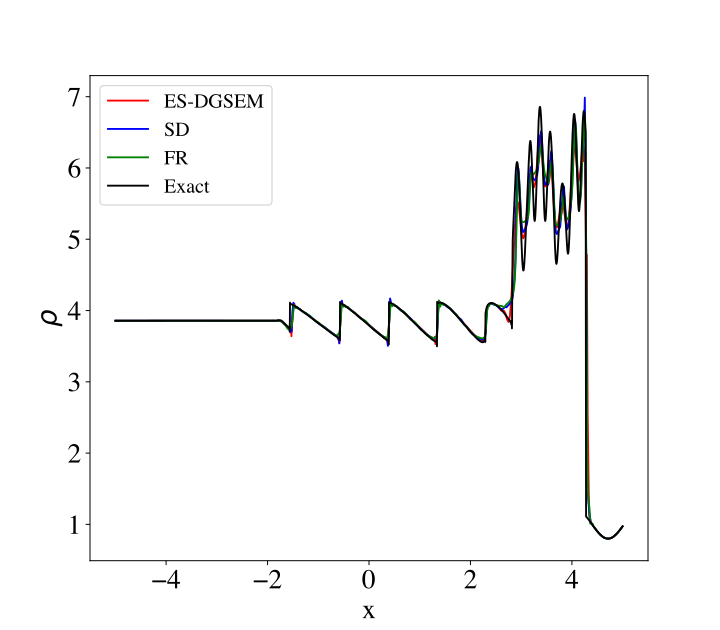}
 &
\includegraphics[width=0.48\textwidth,height=0.38\textwidth]{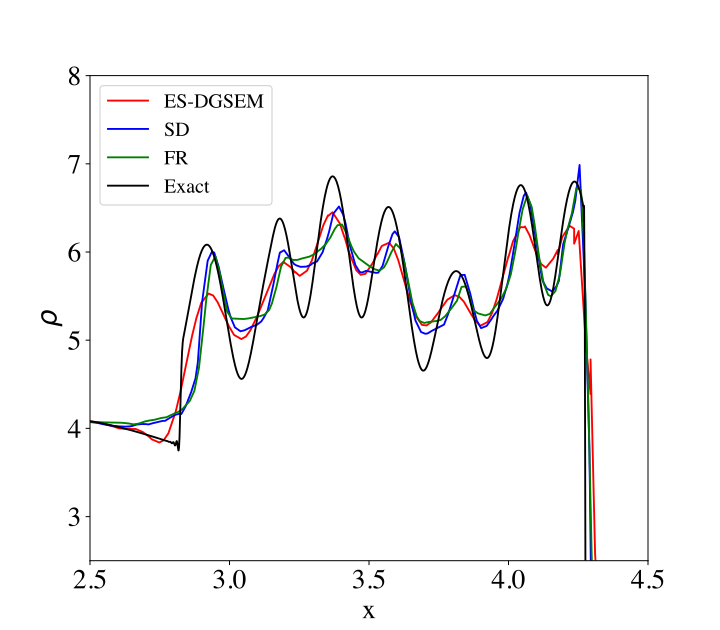}
    
  \\
  (a) Density& (b) Density zoomed
  \\
\includegraphics[width=0.48\textwidth,height=0.38\textwidth]{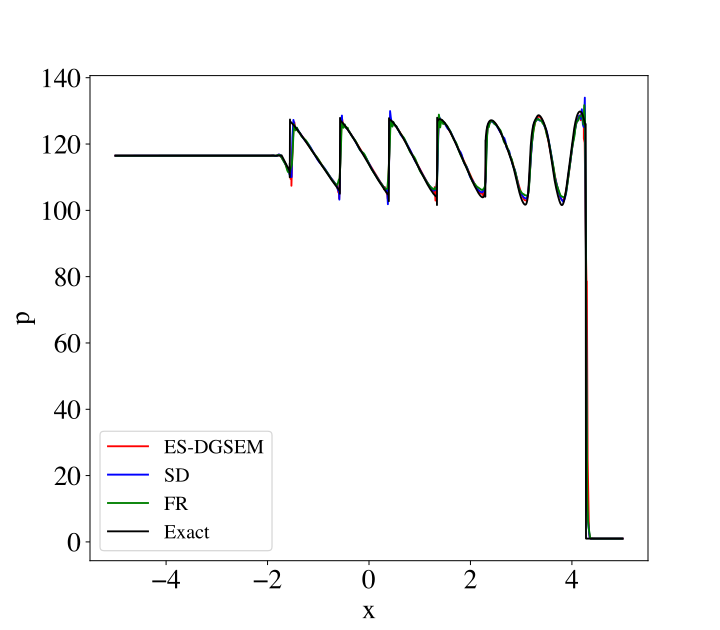}
 &
\includegraphics[width=0.48\textwidth,height=0.38\textwidth]{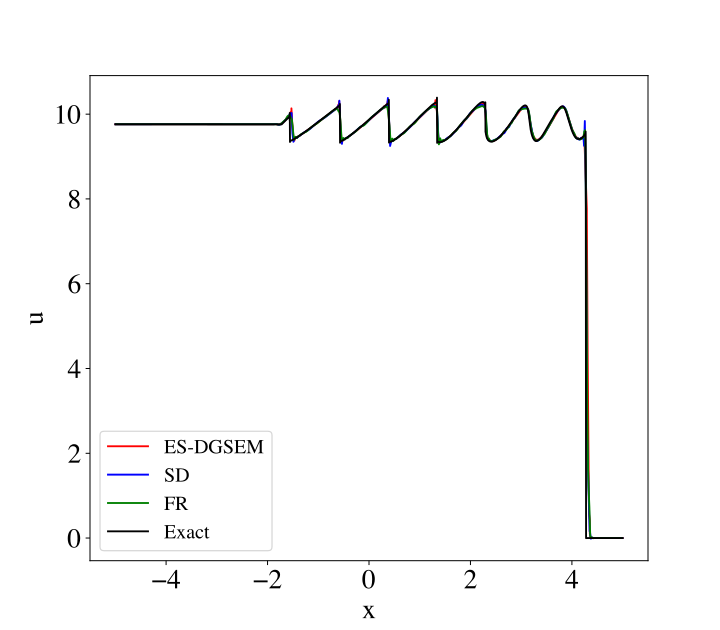}
  \\
  (c) Pressure& (d) Velocity
\end{tabular} 

\caption{Benchmark 2: Results of the modified Shu-Osher shock-density wave interaction test problem. Shown are comparisons between ES-DGSEM (red solid line), SD (blue solid line), and FR (green solid line) methods against the reference (``Exact") results (black solid line) for the benchmark problem. The shock wave speed for the flow is $M_s=10$, and the pressure ratio is 100. The simulation final time is $t=0.7$. A total of $170$ elements are used with the polynomial order of $\mathcal{P}=3$. (a) Density, (b) Density zoomed at the high-frequency region, (c) pressure, and (d) velocity.}

    \label{fig:shuosherm}
  \end{center}
  
\end{figure}

Figures~\ref{fig:shuosherm}(a)-(d) present the modified shock-density wave interaction problem. The FR, SD, and ES-DGSEM results compare the accuracy of capturing high-frequency features that could appear in shock-turbulence interaction scenarios. The stabilization scheme for shock capturing is adaptive modal filtering for FR and SD schemes and high and low order blending technique for the ES-DGSEM. According to Fig.~\ref{fig:shuosherm}(a), all the schemes can successfully resolve all the discontinuous and continuous parts of the solution with slight differences. Looking at Fig.~\ref{fig:shuosherm}(b), the SD and FR schemes fail to capture the dip at $x=2.82$ correctly, and the FR method misses to predict the valley at $x=3.265$ while the ES-DGSEM successfully resolves all the features of the solution. The adaptive filter maintains the numerical entropy at all the solution and flux points above a threshold, which is the minimum numerical entropy within the Voronoi elements (neighboring elements) at the previous Runge Kutta sub-step. This assumption may not correctly operate when there are high-frequency regions besides the shock or contact waves in the domain, where the adaptive filter can be switched on due to the high gradient smooth regions and smear some features of the solution. The selected numerical threshold for entropy $\epsilon=10^{-4}$ must be modified to avoid excessive filtering at high-gradient smooth regions. Therefore, the adaptive modal filter must be tuned according to the problem. However, for the ES-DGSEM scheme the shock capturing is not problem dependent and the tuning parameter is usually kept at $0.5$. The shock wave captured by the SD and FR schemes is sharper than the DGSEM scheme while introducing overshoots and undershoots. The SD undershoots and overshoots are the largest among the three schemes.

%%%%%%%%%%

\subsection{Benchmark 3: Three species Sod problem with simple chemistry}

The three species Sod problem was first presented by Zhang and Shu \cite{zhang2011positivity} to evaluate the positivity preserving technique for the Euler equations with source term due to the chemical reactions. The air is a mixture of three species, including $O$, $O_2$, and $N_2$. The initial conditions are defined as 

\begin{equation}
\left(\rho_O, \rho_{O_2}, \rho_{N_2}\right)=\begin{cases}
\left(5.2518963112572040\times10^{-5},3.748071704863518\times10^{-5}, 2.962489471973072\times10^{-4}\right) & x \le 0.0 \\ 
\left(8.341661837019181\times10^{-8},9.455418692098664\times10^{-11},2.748909430004963\times10^{-7}\right) & x > 0.0, 

\end{cases}
\label{threespecies_den}
\end{equation}

\begin{equation}
\left(u,p\right)=\begin{cases}
\left(0, 1000\right) & x \le 0.0 \\ 
\left(0,1\right) & x > 0.0.

\end{cases}
\label{threespecies_rest}
\end{equation}
The initial temperature is $8,000 $ $K$ in the entire domain. The physical domain is defined as $x \in [-1,1]$. The chemical mechanism includes a single reaction as 

\begin{equation}
O_2+N_2\rightleftharpoons 2O+N_2,
\label{threespecies_rest}
\end{equation}
which introduces a production rate of $\omega$ as defined in \cite{zhang2011positivity} to species transport equations. The production rate of species is implemented for $O$ and $O_2$ species transport equations while $N_2$ acts as a neutral component with zero net production rate. The final simulation time is $t=0.0001$, and the CFL number is 0.1 for all the test cases.

%%%%%%%%%%
\begin{figure}[!htbp]
  \begin{center}
    \begin{tabular}{cc}
     \includegraphics[width=0.48\textwidth,height=0.38\textwidth]{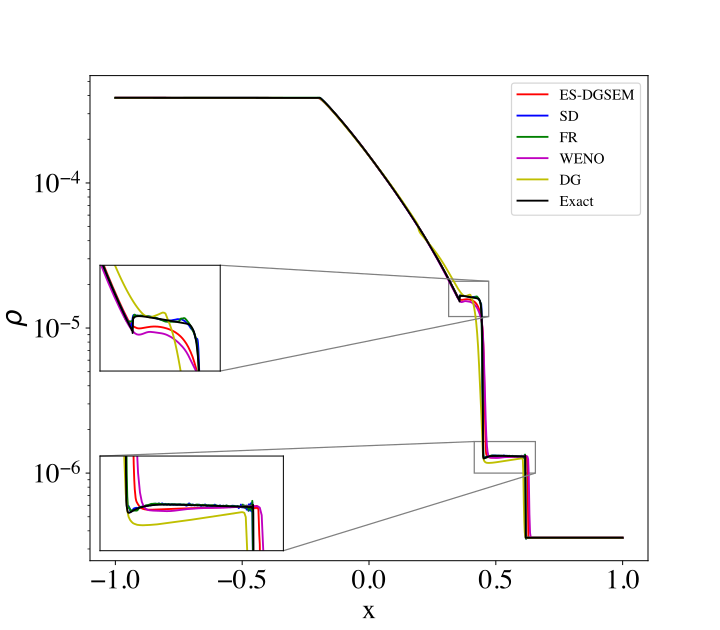}
 &
\includegraphics[width=0.48\textwidth,height=0.38\textwidth]{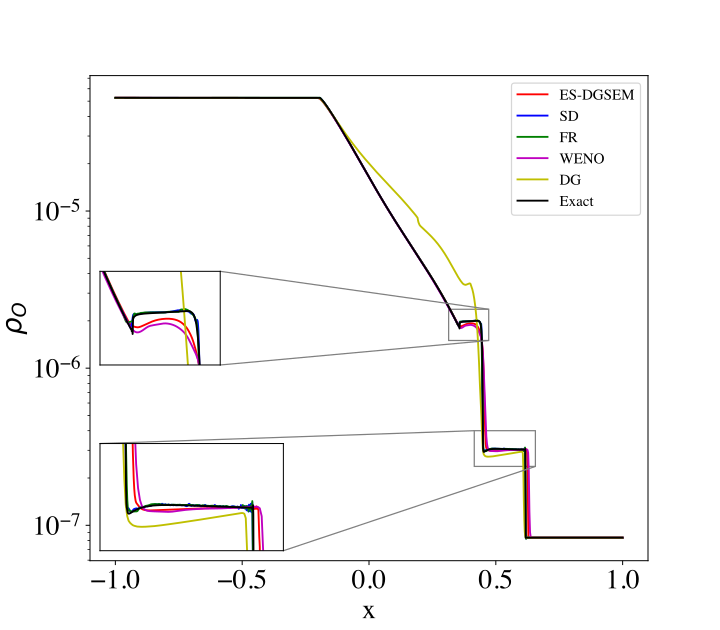}
    
  \\
  (a) Density& (b) O density
  \\
\includegraphics[width=0.48\textwidth,height=0.38\textwidth]{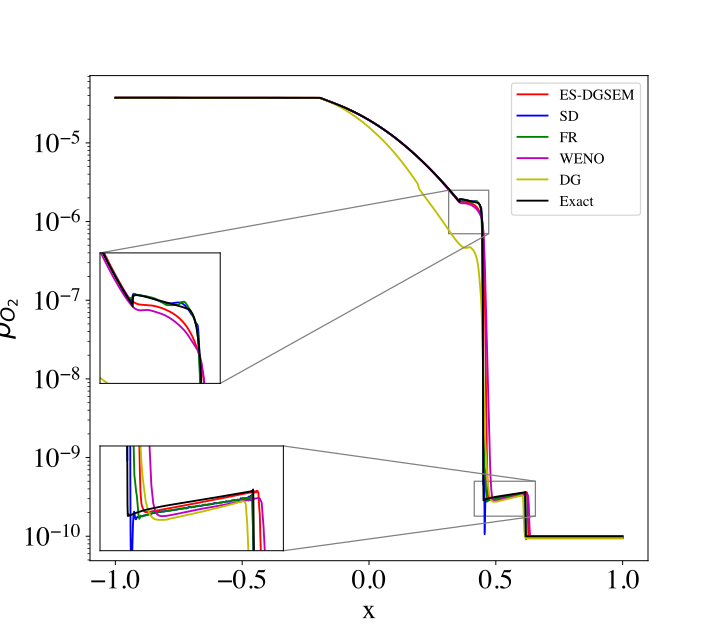}
 &
\includegraphics[width=0.48\textwidth,height=0.38\textwidth]{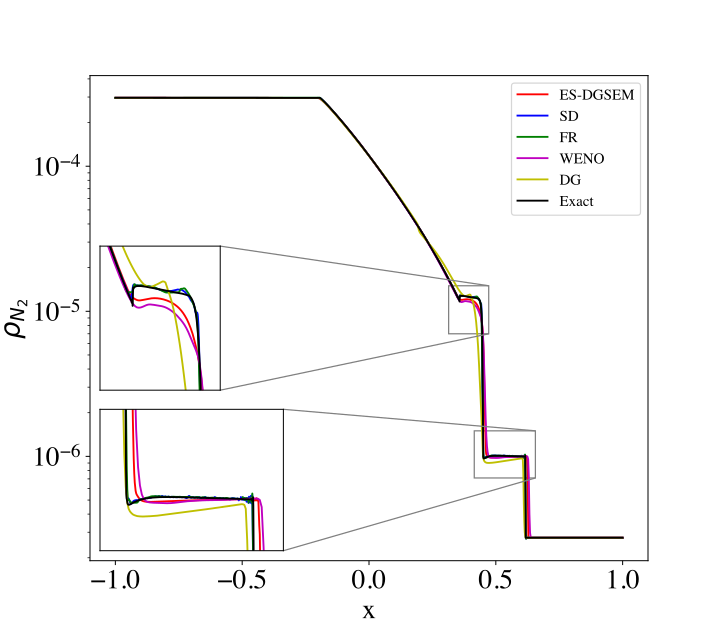}
  \\
  (c)  $O_2$ density& (d) $N_2$ density
 \\
 \includegraphics[width=0.48\textwidth,height=0.38\textwidth]{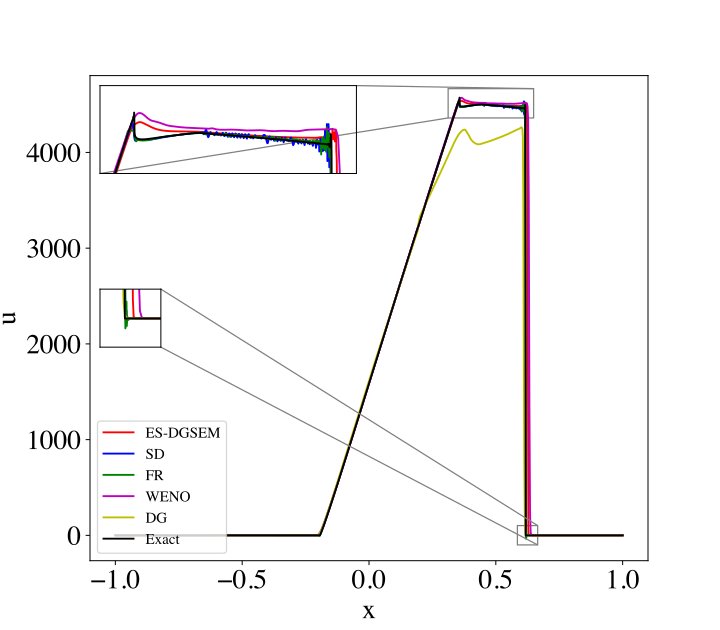}
 &
\includegraphics[width=0.48\textwidth,height=0.38\textwidth]{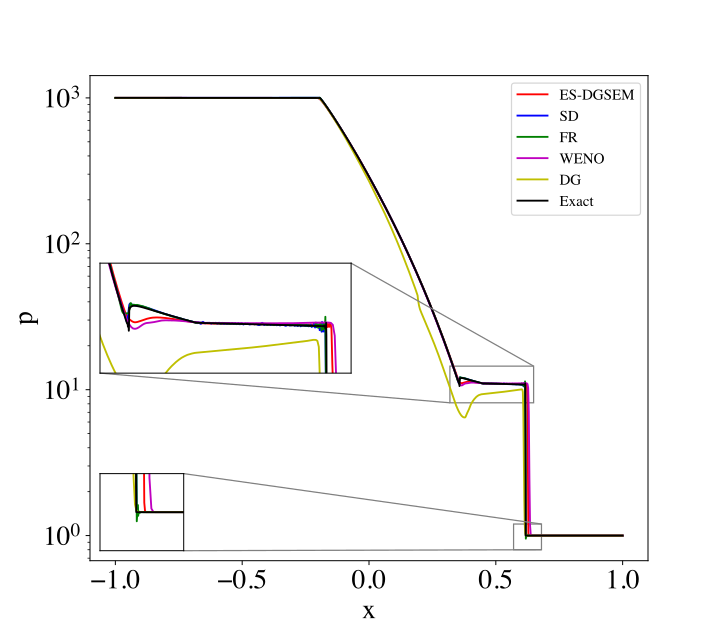}
  \\
  (e) Velocity & (f) Pressure
\end{tabular} 

\caption{Benchmark 3: Results for three species Sod problem with simple chemistry. The gas mixture is modeled with three species $N_2$, $O_2$, and $O$. A single chemical reaction is considered that expresses the dissociation of $O_2$ into $O$. 
For the benchmark problem, five high-order methods are compared. The pressure ratio is $10^3$. A total of 800 elements with $\mathcal{P}=3$ is employed for the ES-DGSEM, SD, and FR approaches.}
    
    \label{fig:threespecies}
  \end{center}
  
\end{figure}

Figures~\ref{fig:threespecies}(a)-(f) compare five high-order schemes in capturing strong shock waves in chemically reactive flows. The DG scheme of Zhang and Shu \cite{zhang2011positivity}, the fifth order finite difference WENO scheme of Zhang and Shu \cite{zhang2012positivity}, FR, SD, and ES-DGSEM schemes are employed to simulate the three species sod problem. The polynomial order for ES-DGSEM, FR, and SD is set at $\mathcal{P}=3$. In contrast, for DG, the polynomial order is $\mathcal{P}=2$, corresponding to the third-order degree of accuracy. The number of elements is 800 for all the schemes except for DG, where we used 1066 elements to have the same number of degrees of freedom as the other spectral element methods. A high-resolution case with $N=4000$ and $\mathcal{P}=3$ is simulated with the ES-DGSEM scheme and considered as an ``exact" (reference) solution. According to Fig.~\ref{fig:threespecies}, all high-order schemes successfully obtain the correct solution except for the DG scheme, which shows a non-physical prediction of species densities, pressure, and velocity. The FR and SD schemes results show a better agreement than WENO and ES-DGSEM schemes while introducing high-frequency oscillations shown in Fig.~\ref{fig:threespecies}(e). The ES-DGSEM scheme shows no oscillations nor the WENO approach while smearing the discontinuities more than the FR and SD methods. Figure~\ref{fig:threespecies}(c) reveals that the SD scheme generates a large undershoot compared to other approaches.

\subsection{Benchmark 4: Hypersonic smooth initial condition problem}
In this section, we demonstrate that the two points' conservative entropy fluxes recover local and global entropy conservation conditions \cite{chan2018discretely}. For this purpose, a test problem is carefully devised to eliminate all the external factors that can destroy entropy conservation. For the hypersonic ES-DGSEM, the interface Riemann solver is replaced by the two points entropy conservative fluxes. The stabilization and positivity-preserving schemes are turned off. The reaction source terms are disabled. The initial conditions are shown in Table.~\ref{tbl:smoothhyper},
\begin{table}[!htbp]
\caption{Constant initial conditions for the smooth initial condition hypersonic test problem}  
\centering
\begin{tabular}{ccc} 
\hline
$p$ ($Pa$)&$\rho$ ($kg/m^3$)&$u$ ($m/s$)\\
\hline
195256.0 & 0.03805 & 11450\\
\hline
\end{tabular}
\label{tbl:smoothhyper}
\end{table}
and the temperature is initialized using a sinusoidal profile defined as 
\begin{equation}
T(x,0)=9000+200\sin(2\pi x).
    \label{temp}
\end{equation}
We then initialized the mass fractions using equilibrium conditions at $T=T(x,0)$ and $p=195256$. The physical domain is defined as $x \in [0,1]$. Periodic boundary conditions are assigned at $x=0$ and $x=1$. Since entropy conservative fluxes are employed for the interior flux points and at the elements' interfaces, the global entropy rate should remain at machine zero over time. 

\begin{figure}[t!]
  \begin{center}

     \includegraphics[width=0.6\textwidth,height=0.35\textwidth]{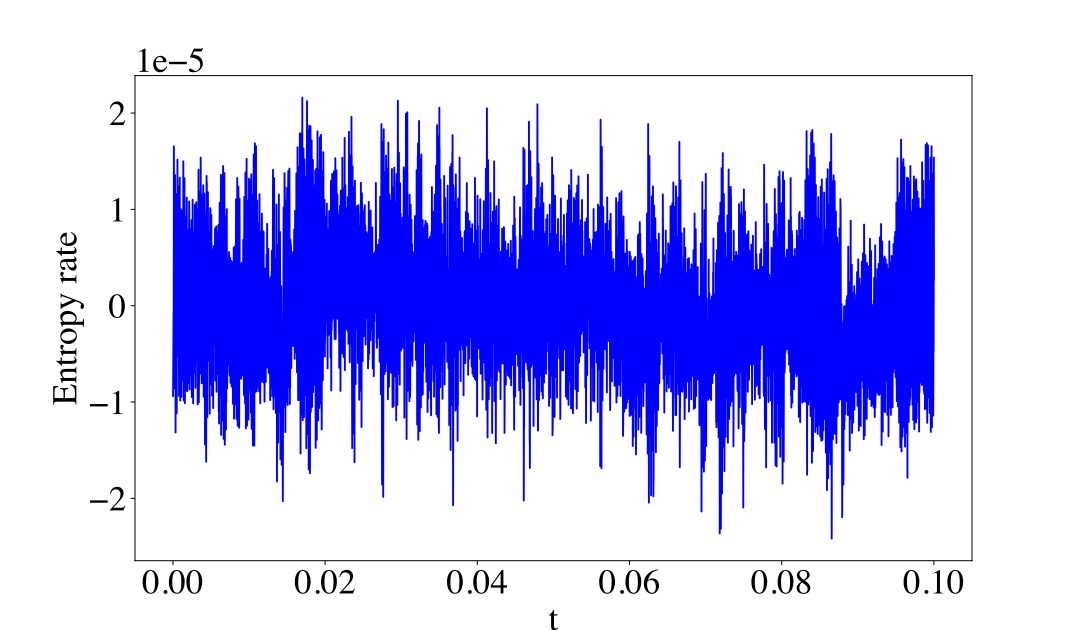}

\caption{Benchmark 4: The global rate of entropy over time for the hypersonic smooth initial condition for Euler equations obtained by the pure entropy conservative EC- DGSEM. A smooth sinusoidal initial condition for temperature is imposed on the domain, and an entropy conservative EC-DGSEM is employed to calculate the solution until $t=0.1$. A total of 170 elements with $\mathcal{P}=3$ are used. The numerical tolerance for inverting the internal energy to obtain the temperature is set at the order of $10^{-7}$.}
    
    \label{fig:smoothhyper}
  \end{center}
  
\end{figure}

\begin{figure}[t!]
  \begin{center}
    \begin{tabular}{ccc}
     \includegraphics[width=0.32\textwidth,height=0.28\textwidth]{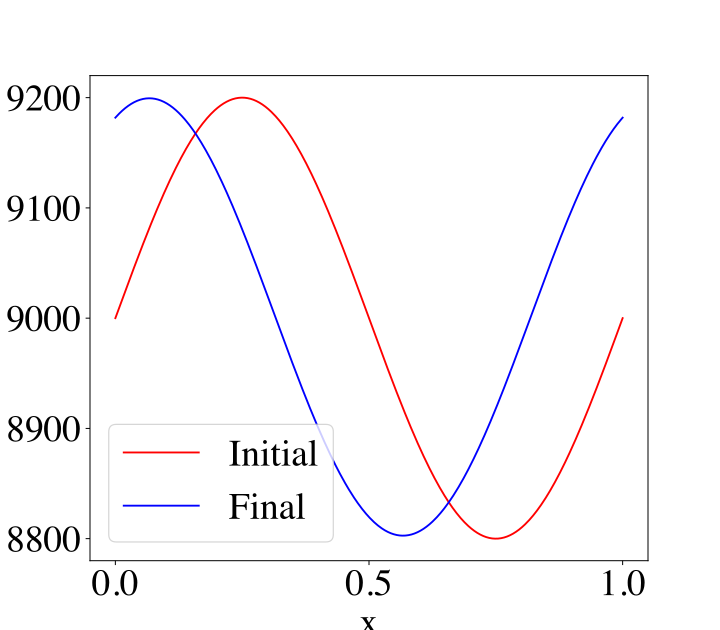}
 &
\includegraphics[width=0.32\textwidth,height=0.28\textwidth]{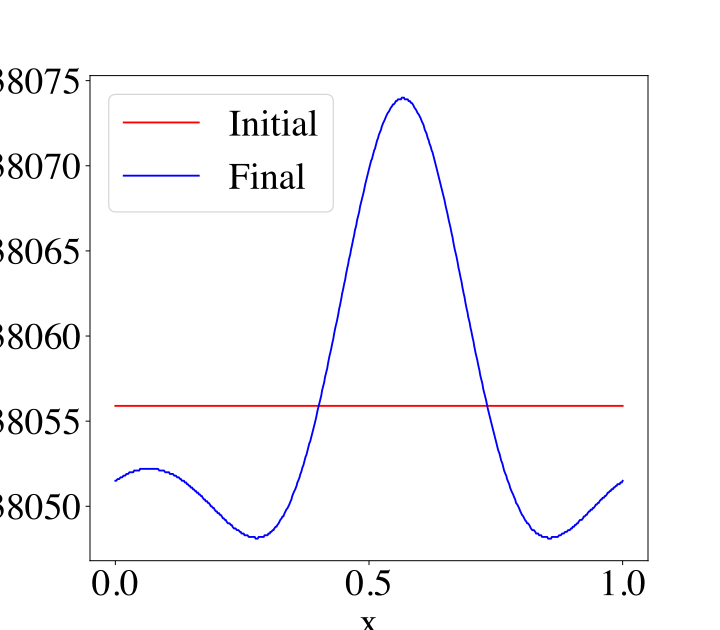}

&
\includegraphics[width=0.32\textwidth,height=0.28\textwidth]{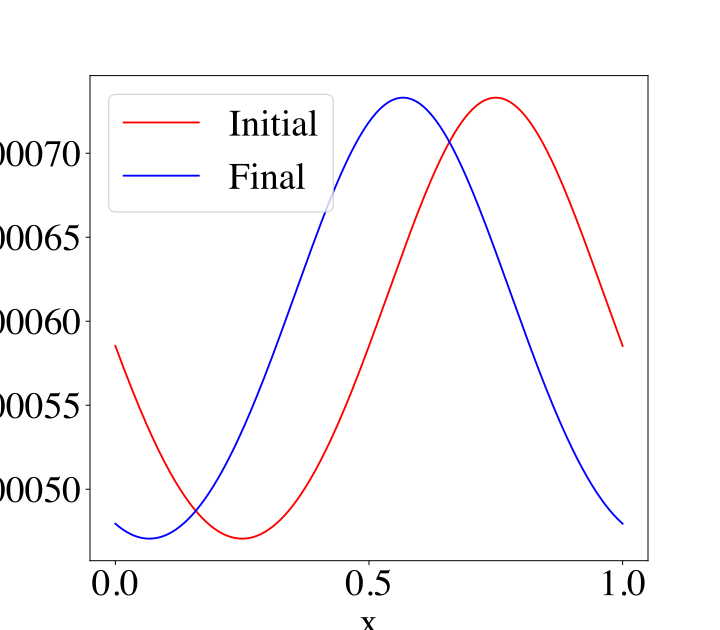}
    
  \\
  (a) Temperature& (b) Density & (c) $Y_{NO}$

\end{tabular} 

\caption{Benchmark 4: Initial and final conditions for the the hypersonic smooth initial condition problem. The blue solid lines indicate the final solution and the red lines denote the initial conditions.}
    
    \label{fig:smoothfield}
  \end{center}
  
\end{figure}

The global entropy rate is determined by multiplying the Eq.~\eqref{strong} without the source and diffusive fluxes into the discrete entropy variables inside each element and calculating the entropy rate of each element using the quadrature rules. We then sum the entropy rate of each element over the entire physical domain to calculate the global entropy rate. The global entropy rate is determined for each time step until $t=0.1$, and the rate of entropy change in time is plotted in Fig.~\ref{fig:smoothhyper}. The initial and final time profiles are also plotted for three variables in Fig.~\ref{fig:smoothfield}. According to Fig.~\ref{fig:smoothhyper}, the rate of entropy change remains within the $10^{-5}$ order of magnitude over time. The value of the global entropy rate is not machine zero; however, it is close to the numerical tolerance defined for inverting the internal energy to the temperature in the Mutation++ library. For calculating the temperature from the internal energy value, the Mutaion++ library uses an iterative loop that uses Newton-Rophson's method to determine temperature until the residual becomes less than a specific numerical tolerance. In the Mutation++ library, this tolerance is in the order of $10^{-7}$, which becomes a limiting factor for the entropy conservation error. We can observe that the global entropy change rate remains close to the numerical tolerance over time. We can thus conclude that the entropy conservative fluxes expressed in Eqs.~\eqref{ECfluxf1}-\eqref{ECfluxf3} retain the global entropy conservation. We also validate that the Tadmor relation is retrieved when the two points' EC fluxes are multiplied into the jump in entropy variables, the product should be equal to the jump in potential function $\psi$.

%%%%%%%%%%

%%%%%%%%%%

\subsection{Benchmark 5: Inviscid hypersonic Sod problem with real chemistry}

\subsubsection{Benchmark 5A}
In this section, we consider the hypersonic Sod problem presented by Grossman and Cinnella \cite{grossman1990flux} with non-equilibrium chemistry to validate the hypersonic 1D ES-DGSEM approach. The ES-DGSEM scheme, along with the high-low order blending technique for advective fluxes, is the most robust high-order method among SD, FR, WENO, and the DG schemes considered in previous benchmark problems. The ES-DGSEM scheme employs the least number of ad-hoc user-defined parameters to resolve the high-pressure ratio shock waves and provides an accurate solution with minimal numerical oscillations. Therefore, we extend  the ES-DGSEM scheme for solving five species of hypersonic flows and use it for the rest of this study. The FR and SD schemes extension for the five-species non-equilibrium sod problem is unstable. The stabilizing method cannot find an optimal filter strength to satisfy the positivity of the five species densities, pressure, and numerical entropy criterion at once.

The non-equilibrium Sod problem consists of the 1D physical domain divided by a membrane into the driver (high pressure) and driven (low pressure) sections. Each section at the beginning is in thermal and chemical equilibrium conditions. Suddenly, the high-pressure and high-temperature driver section contacts the driven section, and shock waves create high-temperature regions, which trigger the finite rate chemical reactions. The dimensional initial conditions are defined as 

\begin{equation}
\left(T(K), u\left(\frac{m}{s}\right), p (Pa)\right)=\begin{cases}
\left(9000, 0,195256\right) & x \le 0.5 \\ 
\left(300, 0, 10000\right) & x > 0.5, 

\end{cases}
\label{grossman}
\end{equation}
for $x \in [0,1]$. Based on the initial pressure and temperature conditions, the equilibrium mass fraction and total density are determined as presented in Table~\ref{tbl:sodgross}. The air composition is assumed to be 79\% nitrogen $N$ and 21\% oxygen $O$. The viscosity, thermal conductivity, and mass diffusion are neglected for this non-equilibrium Sod problem. The reference velocity, temperature, density, and length are considered as $300\, m/s$, $300\, K$, $1 \, kg/s$, and $1 \, m$, respectively. The Eckert number is unity. The solution of the non-equilibrium Sod problem is obtained at a moment when the shock front reaches $x=0.61$, which translates to the non-dimensional time of $t=0.03$. $ N=200$ elements are used to discretize the domain, and the polynomial order is set at $\mathcal{P}=3$. Due to the stiffness of the system of equations with reaction source terms, the CFL number is set at $0.01$ to resolve large production rates due to the high temperature.

\begin{table}[!htbp]
\caption{Benchmark 5A: The equilibrium  total density and mass fractions for driver and driven sections of the non-equilibrium sod problem.}  
\centering
\begin{tabular}{lcccccc} 
\hline
&$\rho$&$Y_N$&$Y_O$&$Y_{NO}$&$Y_{N_2}$&$Y_{O_2}$\\
\hline
$x<0.5$&0.038472&0.725523 &0.232420 &  9.0796e-4&     0.041135 &  1.3042e-05\\
$x>0.5$&0.1156630&7.6357e-80 &  4.0011e-41 & 2.6973e-16   &  0.767082& 0.232918\\
\hline
\end{tabular}
\label{tbl:sodgross}
\end{table}

%%%%%%%%%%
\begin{figure}[t!]
  \begin{center}
    \begin{tabular}{ccc}
     \includegraphics[width=0.32\textwidth,height=0.28\textwidth]{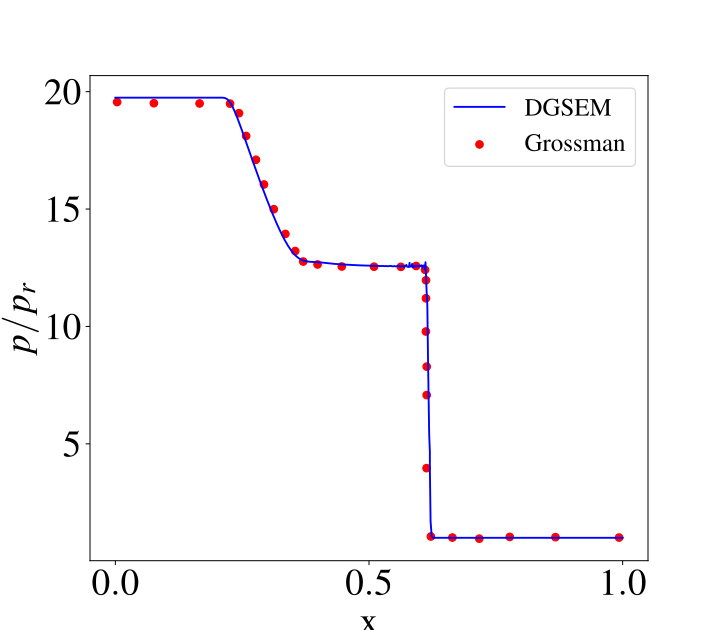}
 &
\includegraphics[width=0.32\textwidth,height=0.28\textwidth]{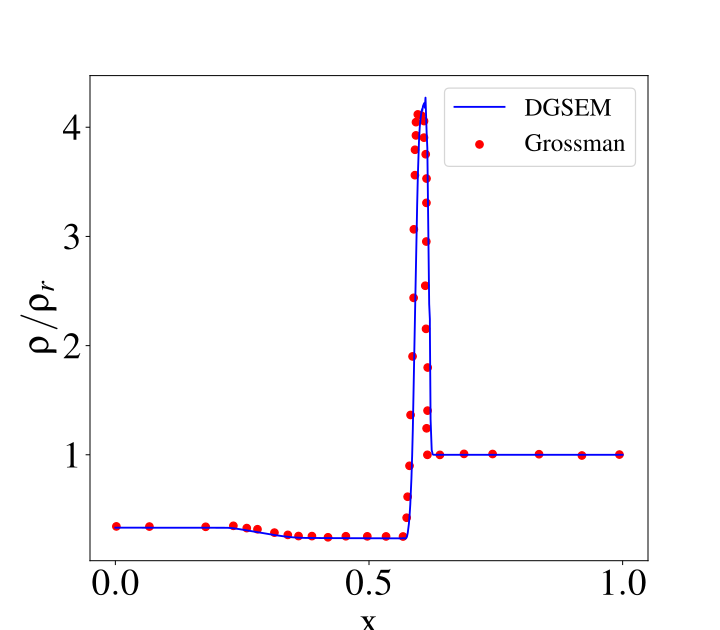}

&
\includegraphics[width=0.32\textwidth,height=0.28\textwidth]{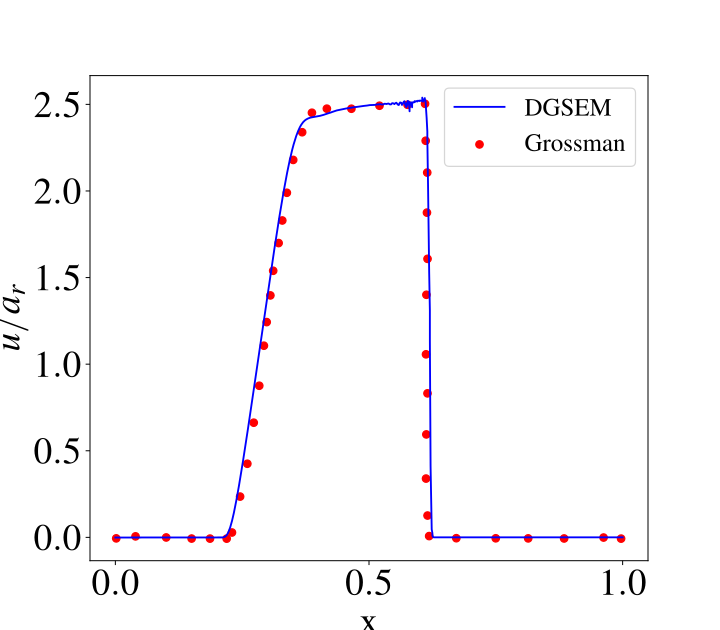}
    
  \\
  (a) Pressure ratio& (b) Density ratio& (c) Velocity ratio
  \\
\includegraphics[width=0.32\textwidth,height=0.28\textwidth]{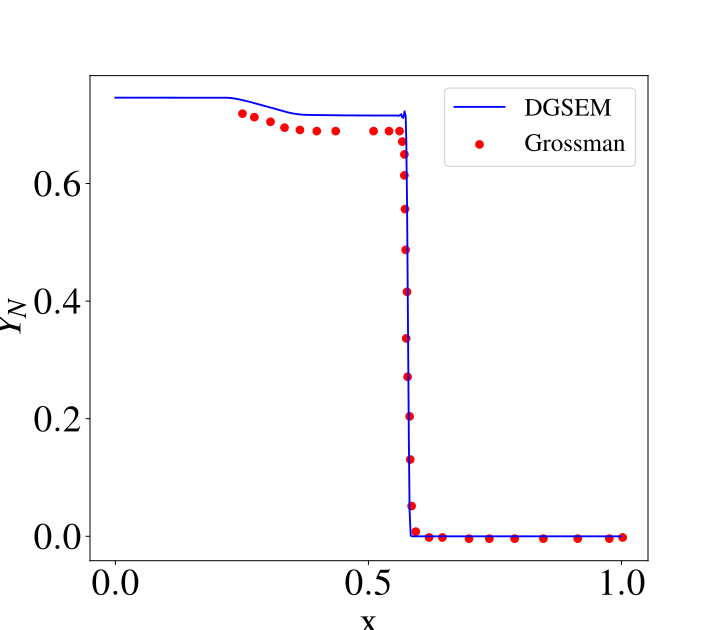}
&
 \includegraphics[width=0.32\textwidth,height=0.28\textwidth]{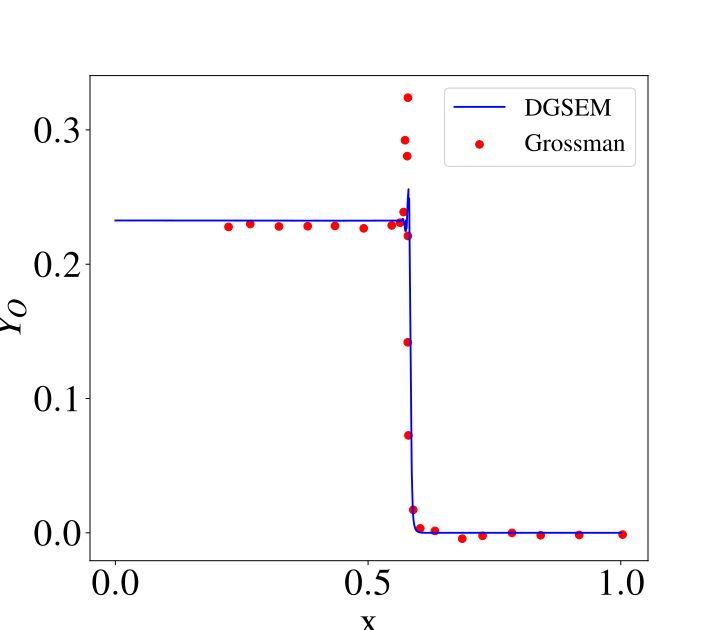}
&
\includegraphics[width=0.32\textwidth,height=0.28\textwidth]{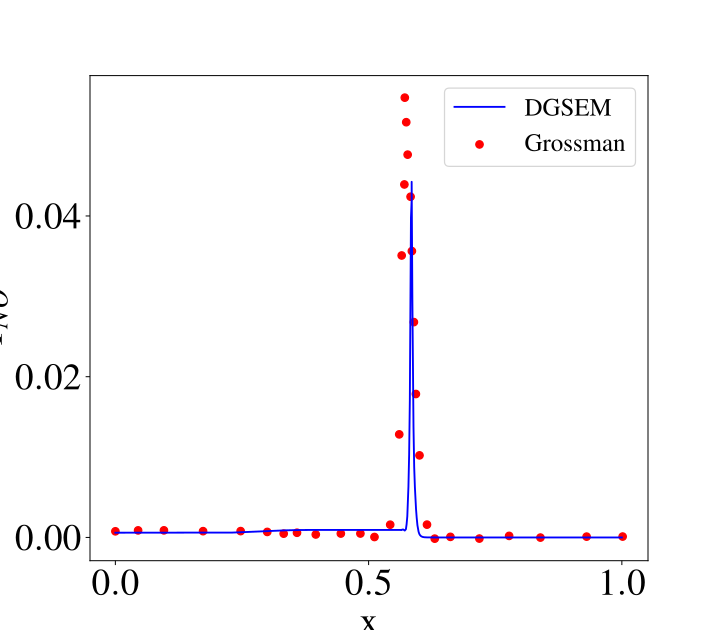}
  \\
  (d) $N$ mass fraction& (e) $O$ mass fraction &(f) $NO$ mass fraction
 \\
 \includegraphics[width=0.32\textwidth,height=0.28\textwidth]{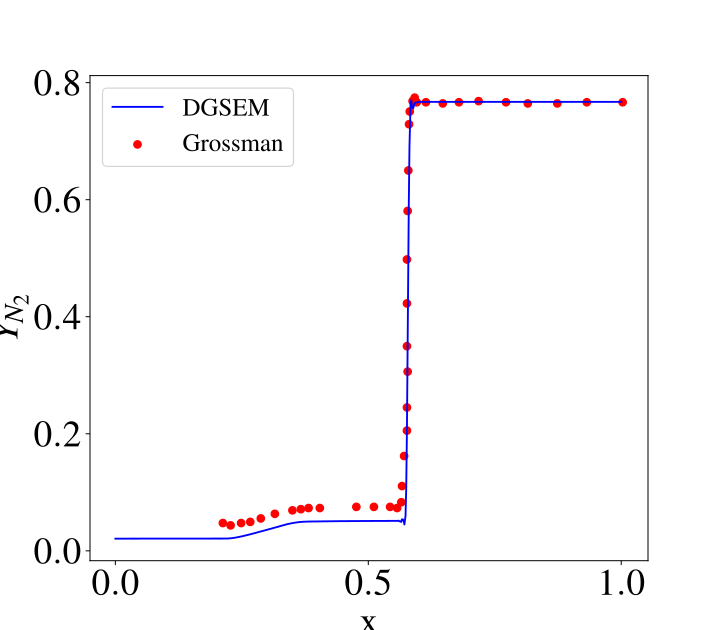}
&
 \includegraphics[width=0.32\textwidth,height=0.28\textwidth]{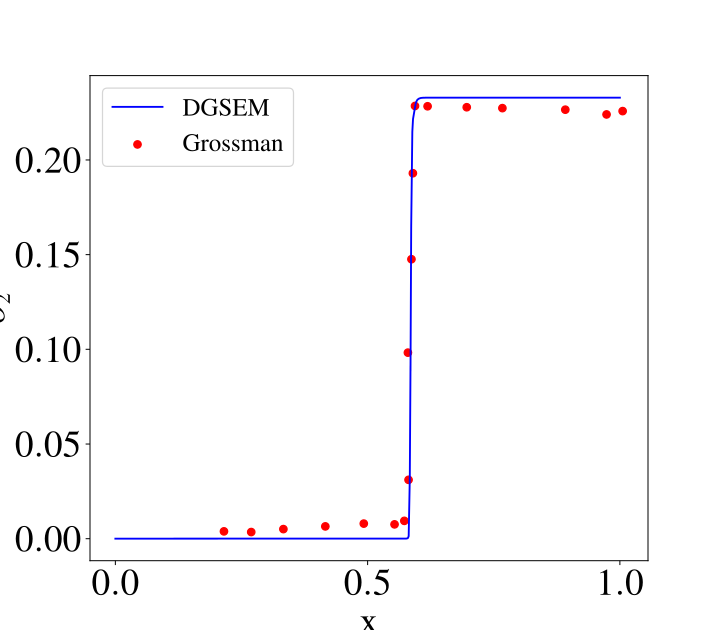}
&
\\
  (g) $N_2$ mass fraction& (h) $O_2$ mass fraction &
\end{tabular} 

\caption{Benchmark 5A: Inviscid hypersonic Sod problem with real chemistry. The problem is first presented in the work of Grossman and Cinnella \cite{grossman1990flux}. The red circles are results from the reference \cite{grossman1990flux} and the blue solid lines show results obtained from the ES-DGSEM scheme. 200 elements with $\mathcal{P}=3$ are employed.}
    
    \label{fig:grossman}
  \end{center}
  
\end{figure}

Figure~\ref{fig:grossman} presents the validation of the hypersonic ES-DGSEM approach for the non-equilibrium Sod problem. In Fig.~\ref{fig:grossman}(a)-(c), the subscript $r$ refers to the initial values of the driven section. The term $a_r$ indicates the speed of sound for the initial conditions at the driven section. All the parameters match the reference results. According to Fig.~\ref{fig:grossman}(b), (e), and (f), all the steep gradient features of the solution are accurately captured by the high-order ES-DGSEM scheme.

%%%%%%%%%%
\begin{figure}[t!]
  \begin{center}
    \begin{tabular}{ccc}
     \includegraphics[width=0.32\textwidth,height=0.28\textwidth]{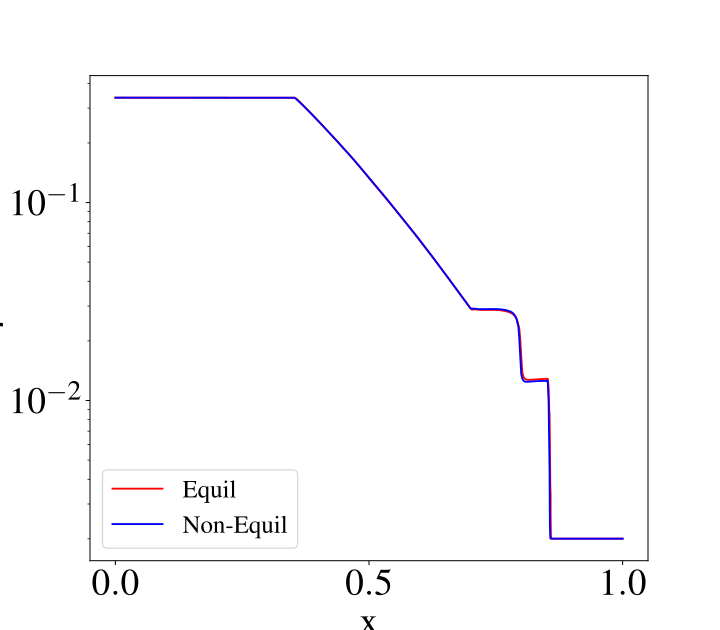}
 &
\includegraphics[width=0.32\textwidth,height=0.28\textwidth]{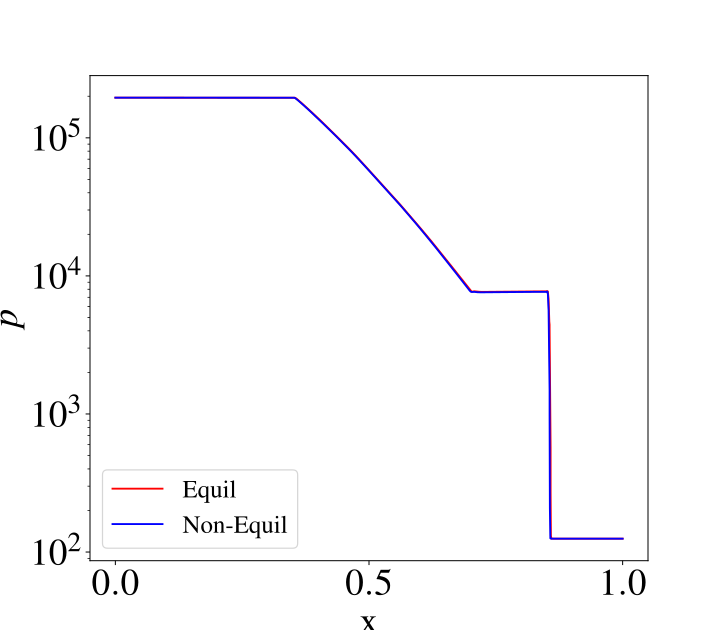}

&
\includegraphics[width=0.32\textwidth,height=0.28\textwidth]{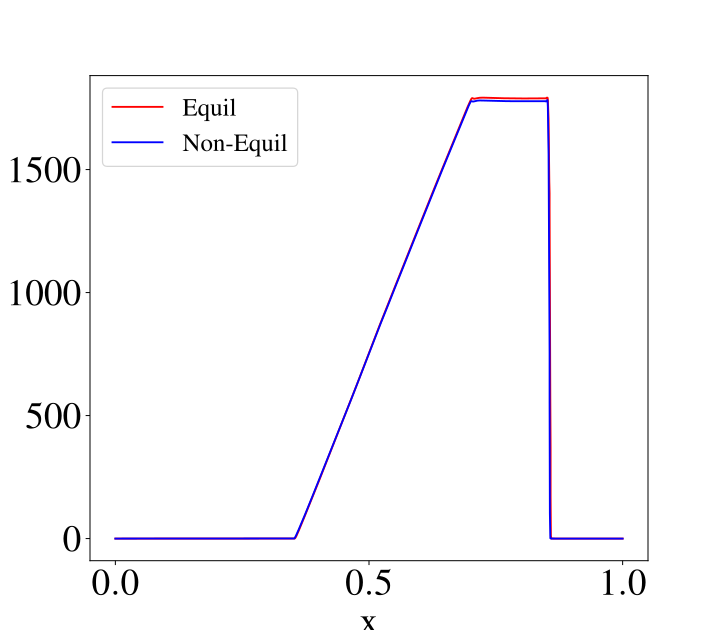}
    
  \\
  (a) Density $T_l=2000\, K$& (b) Pressure $T_l=2000\, K$  & (c) Velocity $T_l=2000\, K$
  \\
\includegraphics[width=0.32\textwidth,height=0.28\textwidth]{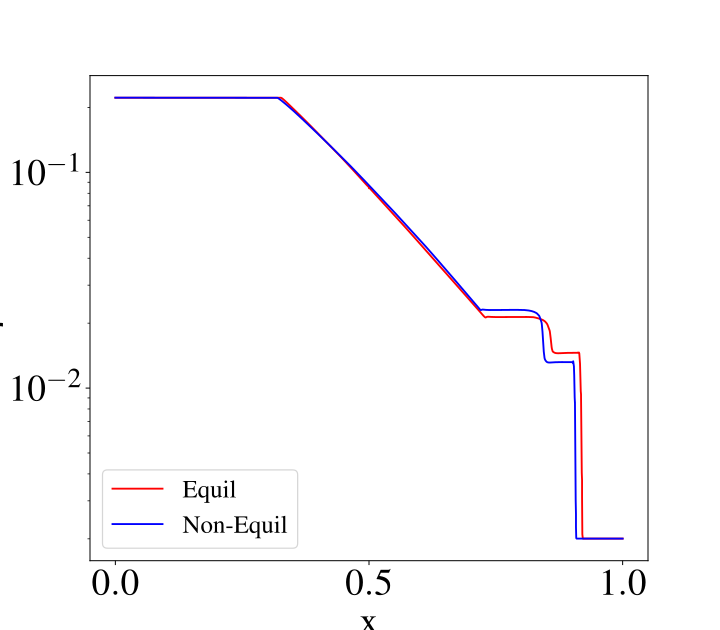}
&
 \includegraphics[width=0.32\textwidth,height=0.28\textwidth]{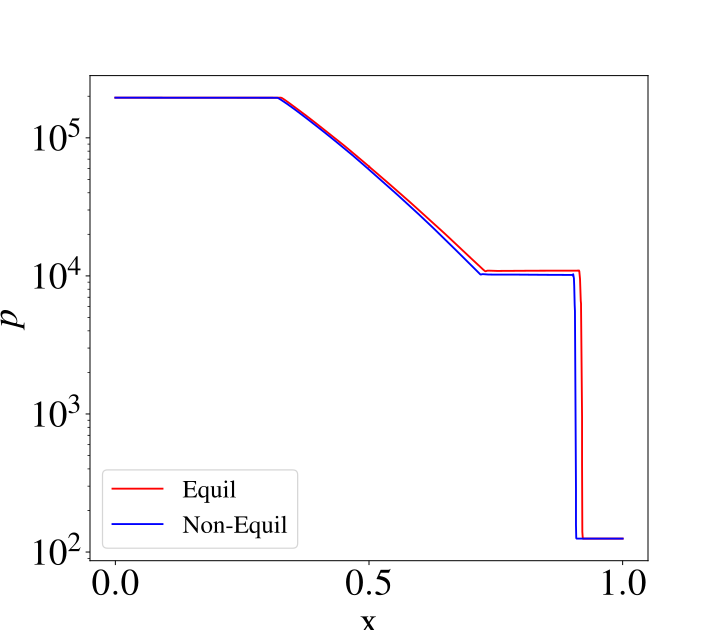}
&
\includegraphics[width=0.32\textwidth,height=0.28\textwidth]{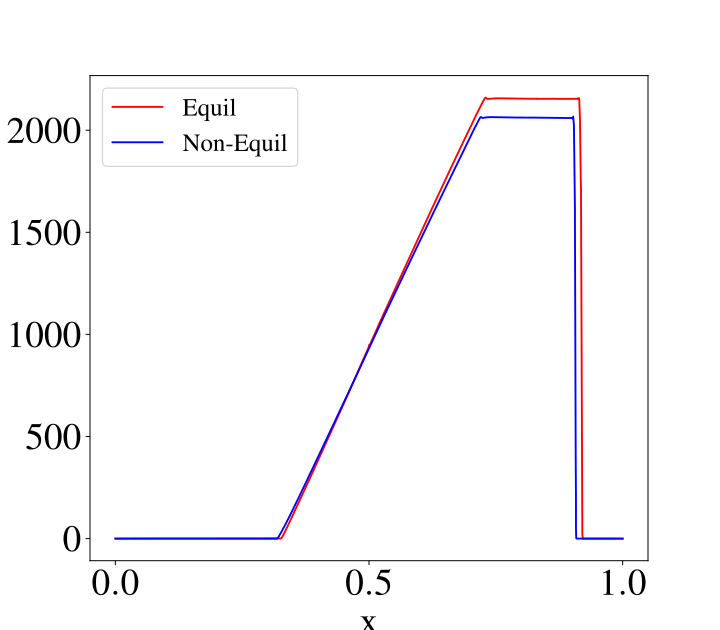}
  \\
  (d) Density $T_l=3000\, K$& (e) Pressure $T_l=3000\, K$  & (f) Velocity $T_l=3000\, K$
 \\
 \includegraphics[width=0.32\textwidth,height=0.28\textwidth]{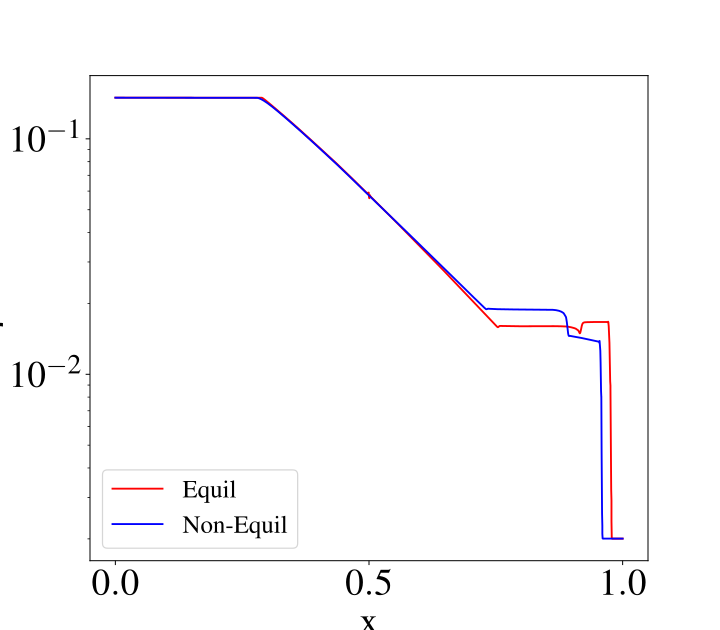}
&
 \includegraphics[width=0.32\textwidth,height=0.28\textwidth]{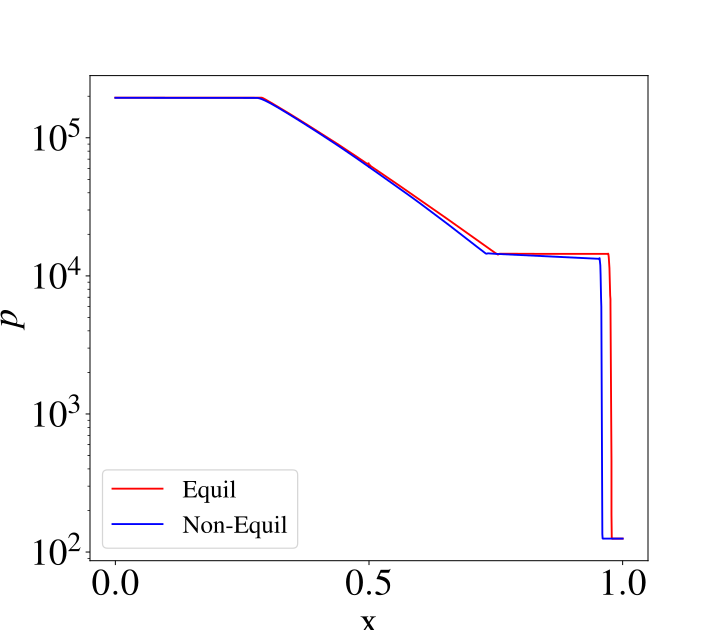}
&
 \includegraphics[width=0.32\textwidth,height=0.28\textwidth]{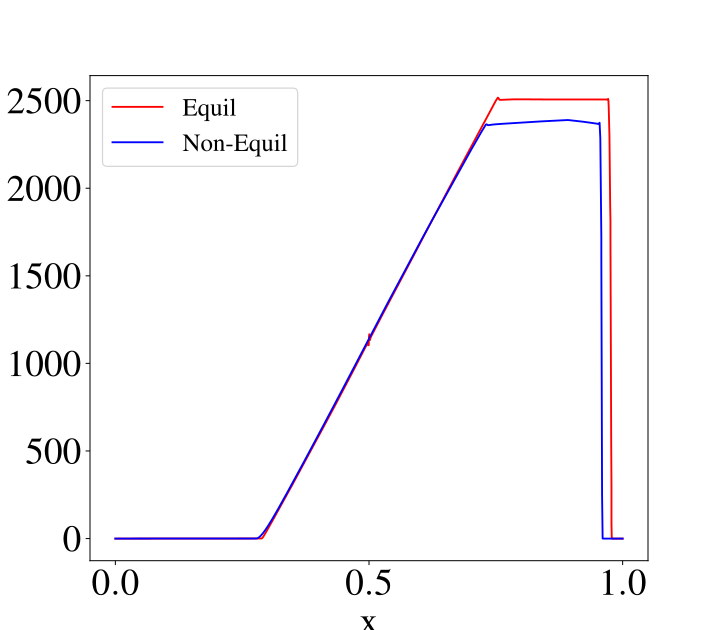}
\\
  (g) Density $T_l=4000\, K$& (h) Pressure $T_l=4000\, K$  & (i) Velocity $T_l=4000\, K$
\end{tabular} 

\caption{Benchmark 5B: A comparison of the real chemistry vs ideal chemistry for a high pressure ratio (1500) Sod problem. The results obtained using equilibrium conditions are shown with red solid lines, while the blue solid lines represent the non-equilibrium chemistry.}
    
    \label{fig:realvsideal}
  \end{center}
  
\end{figure}

\subsubsection{Benchmark 5B}
Next, we select another Sod problem to investigate the effect of real and ideal chemistry when the chemical reaction rates are assumed to be infinity, which correspond to chemical equilibrium conditions. In contrast, real chemistry indicates that chemical reaction evolution occurs at a finite rate. For hypersonic flows, the characteristic time scale decreases to the chemical reaction time scale. This condition renders the assumption of equilibrium condition incorrect. This section shows the conditions under which the equilibrium assumption fails in the following Sod problem. The primitive variables are initialized as 

\begin{equation}
\left(T(K), u\left(\frac{m}{s}\right), p (Pa)\right)=\begin{cases}
\left(T_l, 0,195256\right) & x \le 0.5 \\ 
\left(216.48, 0, 125.07\right) & x > 0.5, 

\end{cases}
\label{equilvsnonequil}
\end{equation}
where the initial condition for $x>0.5$ is the atmospheric condition at the $44\,km$ altitude. The initial temperature for $x<0.5$ ($T_l$) is considered a free parameter. The $T_l$ is increased while keeping the pressure ratio constant at $p_r/p_l=1,561$.

Figure~\ref{fig:realvsideal} presents a comparison between using the equilibrium and non-equilibrium conditions for modeling chemical reactions corresponding to air dissociation. In Fig.~\ref{fig:realvsideal}(a)-(c), the equilibrium assumption is valid since the advection time scale is larger than the chemical reaction time scale. Therefore, the equilibrium condition that assumes infinite reaction rates is valid, and the non-equilibrium results match the results obtained using the equilibrium chemistry. At $T_l=3000\, K$, the equilibrium and non-equilibrium results start to deviate from each other as the advection velocity increases leading to a smaller advection time scale. In Fig.~\ref{fig:realvsideal}(f), we observe that the velocity of the fluid behind the shock front is over-estimated by the equilibrium assumption. The discrepancy between equilibrium and non-equilibrium results is amplified by increasing $T_l=4000\,K$. At this temperature, the advection velocity increases to the point that the convective time scale matches the reaction time scale, violating the equilibrium assumption.

\subsection{Benchmark 6: Steady state viscous flow behind a hypersonic shock wave}
Finally, the hypersonic viscous test case of Marxen \emph{et al.}~\cite{marxen2013method} is considered to validate the implementation of the transport coefficients and species production rates. The benchmark problem features a flow field behind a strong shock wave formed inside a free stream with $M_\infty=10$. The physical domain is defined as a 1D interval with constant initial profiles for temperature, velocity, and five species densities. The initial conditions for the mass fractions, total density, and temperature are shown in Table.~\ref{tbl:marxen}. The initial mass fractions are determined assuming equilibrium condition at $T_\infty=350\textrm{ }K$ and $P_\infty=35.9593\textrm{ }Pa$. The reference values are also calculated using the equilibrium assumption at $T_\infty$ and $P_\infty$ and shown in Table~\ref{tbl:marxeninit}. The non-dimensional $\rho u$ variable is initialized at $9.25$. The 1D interval is defined as $x\in[0,0.02]$ and the simulation is continued until $t_{final}=0.03$ when the steady state is reached. 

\begin{table}[t!]
\caption{Benchmark 6: Initial conditions for the steady state solution behind the shock wave.}  
\centering
\begin{tabular}{ccccccc} 
\hline
T (K)&$\rho$ ($\textrm{kg}/\textrm{m}^3$)&$Y_N$&$Y_O$&$Y_{NO}$&$Y_{N_2}$&$Y_{O_2}$\\
\hline
5918.8677  &0.00255537 &0.0 &0.0 &  0.0&     0.767082 &  0.232918\\
\hline
\end{tabular}
\label{tbl:marxen}
\end{table}

\begin{table}[!htbp]
\caption{Benchmark 6: Reference conditions for the 1D steady state solution behind the shock wave.}  
\centering
\begin{tabular}{ccccccccccc} 
\hline
$T_f$ (K)&$\rho_f$ ($\textrm{kg}/\textrm{m}^3$)&$L_f$ ($m$)&$U_f$ ($m/s$)&$\mu^*_f$ ($kg/ms$)&$\kappa^*_f$ ($W/mK$)&$\gamma_f$&$\textrm{Re}_\infty$&$\textrm{Pr}_\infty$&$\textrm{Ec}_\infty$\\
\hline
139.013  &3.565e-04&1.60312 &375.407 &2.145e-05 &  0.0315333&     1.397 &  10000&0.69&1\\
\hline
\end{tabular}
\label{tbl:marxeninit}
\end{table}
Marxen \emph{et al.}~\cite{marxen2013method} applied Dirichlet boundary condition at the inflow ($x=0.0$) and the outflow conditions are enforced using a penalty source term applied on a sponge region defined as $x\in[0.01394,0.02]$. For the DGSEM scheme, the boundary conditions for the advective fluxes are enforced weakly through the approximate Riemann solver. Therefore, we define a sponge region at the inflow to mimic the Dirichlet condition of the finite difference code in \cite{marxen2013method}. We define a Lagrange multiplier to enforce desired values for the conservative variables at the boundaries. We modify the source term of the equations as 

\begin{equation}
S_i=S_i-\sigma \left(U_i-U_{i,0}\right)\quad i=1,\cdots,N_s+2,
    \label{source_term}
\end{equation}
where $S_i$ is the source term for the $i^\textrm{th}$ transport equation. Subscript $0$ indicates the initial values of the conservative variables calculated with the initial conditions defined in Table.~\ref{tbl:marxen}. The source terms modifications are applied to all the transport equations. The coefficient $\sigma$ is determined as 

\begin{equation}
\sigma=A_s\left(\frac{x-x_{beg}}{x_{end}-x_{beg}}\right)^{N_s},
    \label{sigmadef}
\end{equation}
where subscripts $end$ and $beg$ refer to the x coordinate of the sponge region's beginning and end. At the outflow sponge region, $A_s=960$ and $N_3=3$ similar to the work of Marxen \emph{et al.}~\cite{marxen2013method}. For the inflow, the sponge region is defined as $x\in[-10^{-4},10^{4}]$, $A_s=-10^7$ and $N=3$. In the inflow sponge region, we employed $5$ elements, and for the rest of the domain, $40$ elements are used and the polynomial order is set at $\mathcal{P}=3$. For this test case, the mass diffusion fluxes are set at zero similar to Marxen \emph{et al.} work \cite{marxen2013method}.

%%%%%%%%%%
\begin{figure}[t!]
  \begin{center}
    \begin{tabular}{ccc}
     \includegraphics[width=0.32\textwidth,height=0.28\textwidth]{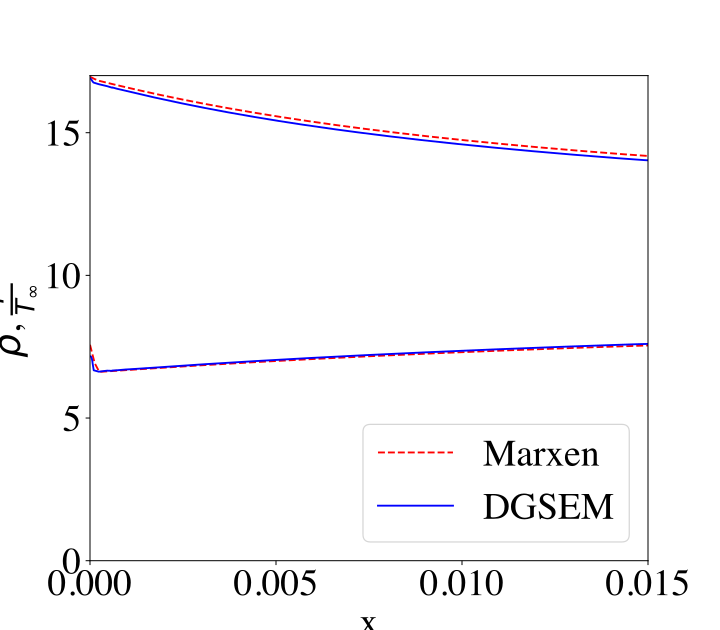}
 &
\includegraphics[width=0.32\textwidth,height=0.28\textwidth]{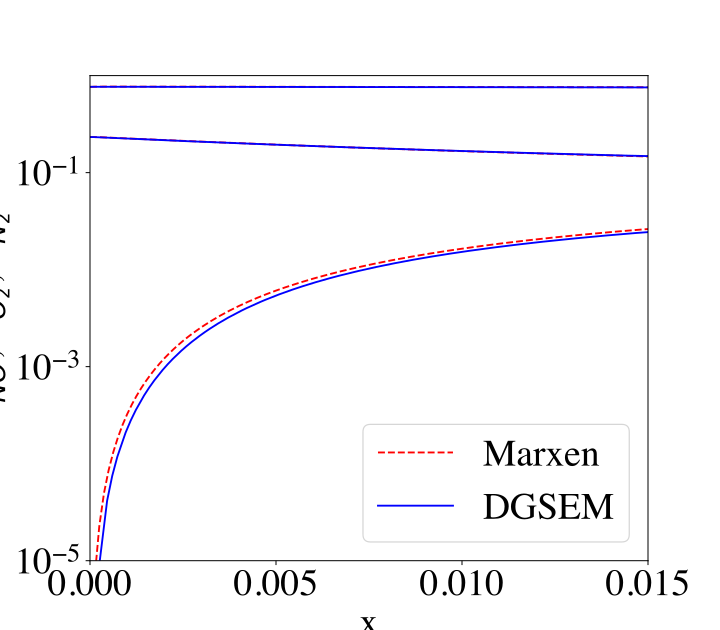}

&
\includegraphics[width=0.32\textwidth,height=0.28\textwidth]{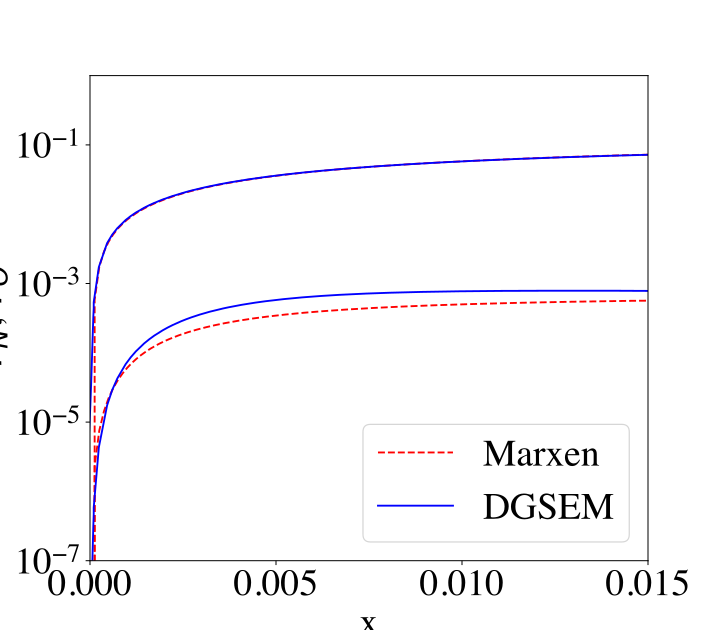}
    
  \\
  (a) $\rho$ and $\frac{T}{T_\infty}$& (b) $Y_{N_2}$, $Y_{O_2}$, and $Y_{NO}$  & (c) $Y_{O}$ and $Y_{N}$

\end{tabular} 

\caption{Benchmark 6: A comparison of the steady state viscous flow behind a hypersonic shock wave results obtained by the viscous version of DGSEM and Marxen \emph{et al.}~\cite{marxen2013method} finite difference scheme. Sponge regions are imposed at the inflow to mimic the Dirichlet boundary conditions and outflow conditions. A total of 45 elements with $\mathcal{P}=3$ are employed to calculate the steady state solution.}
    
    \label{fig:marxen}
  \end{center}
  
\end{figure}
A comparison between the viscous version of DGSEM scheme and Marxen \emph{et al.}~\cite{marxen2013method} finite difference approach is shown in Fig.~\ref{fig:marxen}(a)-(c). The results show a good agreement with the reference results except for a small discrepancy in the $N$ and $NO$ mass fraction profiles. The sources of the discrepancy can be attributed to two reasons. 
In the finite difference code, the boundary condition is directly set as a Dirichlet condition at the inflow boundary. In contrast, for the DGSEM code, the conditions are applied weakly through the Riemann solver for advective fluxes. Second, the finite difference code does not recover the mass conservation, which would ensure the summation of all the mass fractions to be unity. The summation of all the five species mass fractions from Marxen's results adds up to 1.0035 at $x=0.01$ for the steady-state solution. The amount of error in summing all the mass fractions in Marxen's results is in the same order as the difference in the mass fractions comparison. Therefore, the results of the viscous DGSEM scheme are likely more accurate since it preserves mass conservation and enforces unity for the summation of all the species' mass fractions. 

\section{Summary}
In this study, we sought to identify a high-order scheme that could simulate flows with strong shock waves close to vacuum conditions. We began by examining single-species test cases without real chemistry effects. Various existing high-order methods, such as FR, SD, WENO, DG, and DGSEM were considered as well as state-of-the-art stabilization schemes and positivity-preserving techniques for simulating strong shock waves. First, we compared the SD and FR schemes coupled with adaptive entropy modal filtering for positivity preserving and stabilization and entropy stable DGSEM (ES-DGSEM) scheme to simulate the Leblanc problem. With the ES-DGSEM scheme we obtained results with the most negligible amplitude of oscillations compared to the other spectral element methods. We then evaluated the three schemes by simulating the modified shock-density wave interaction problems, where the shock wave Mach number is 10, similar to the hypersonic regimes. The ES-DGSEM scheme can capture smooth high gradient flow conditions and shock discontinuities simultaneously without losing its high accuracy in the smooth regions of the solution. 

We then employed various high-order schemes to simulate the Sod problem with simplified chemistry. The ES-DGSEM scheme reduced oscillations in the vicinity of discontinuities compared to FR and SD schemes. The modal DG obtained inaccurate results, while the WENO approach had  comparable accuracy with the ES-DGSEM scheme. After comparing the high-order methods using relevant benchmark problems, we extended the ES-DGSEM to hypersonic flows with real chemistry. The reason is that the DGSEM scheme can readily scale to large parallel computations compared to the high-order WENO schemes. Also, the FR and SD schemes with adaptive modal filtering are unstable in simulating five species gas mixtures.

The current study presents a high-order DGSEM framework to solve hypersonic flows governing equations with real chemistry models. An entropy stable scheme was first constructed for the hypersonic Euler equation with the five-species gas mixture model. The standard DGSEM scheme was reformulated into flux differencing representation; then, the two-point entropy conservative fluxes were derived to reconstruct a high-order entropy conservative framework. The blending shock-capturing technique was integrated into the entropy conservative method to achieve entropy stability. A positivity-preserving technique was also coupled with the ES-DGSEM to maintain the species' densities and pressure within the physical bounds. A novel formulation for the hypersonic Euler equation was also derived that retrieves the local and global entropy conservation up to the numerical precision enforced by the tolerance of inverting internal energy to the temperature. The entropy-stable DGSEM Euler solver was employed to solve a non-equilibrium Sod problem with stiff reaction source terms. The entropy stable scheme showed good agreement with the reference results and generated a minimal amount of oscillations due to the discontinuities in the solution. The effect of considering real chemistry versus ideal chemistry was also investigated  by considering a hypersonic Sod problem. The results demonstrated that as the driver gas temperature increased above a threshold, the ideal chemistry modeling results deviated drastically from the real chemistry modeling. Therefore, real chemistry modeling is essential for correctly simulating hypersonic flows.

The ES-DGSEM Euler solver was extended to the solution of the reactive Navier-Stokes equations. The diffusive fluxes were discretized using the BR1 approach. The reactive Navier-Stokes solver was then used to simulate a hypersonic viscous 1D steady-state solution behind a strong shock wave. The gas thermodynamic properties, transport coefficients, and chemical reaction source terms were all calculated using the Mutation++ library. The viscous test case verified the correct calculation of the species source terms along with the transport coefficients by coupling the Mutation++ library with the ES-DGSEM solver. In the future, the 1D hypersonic DGSEM scheme can be extended to 2D and 3D dimensions since the ES-DGSEM has been successfully extended to multiple dimensions for non-reactive flows.

\bibliographystyle{unsrt}
\bibliography{cfd} 

\end{document}